\documentclass{new_tlp}
\usepackage[linesnumbered,ruled,vlined]{algorithm2e}
\usepackage{graphicx}
\usepackage{amsmath}
\usepackage{pxfonts}
\usepackage{tikz}
\usepackage{caption}

\makeatletter
\AtBeginDocument{%
  \def\doi#1{\url{https://doi.org/#1}}}
\makeatother
\usepackage{hyperref}

\SetKwInput{KwInput}{Input}  
\SetKwInput{KwOutput}{Output}  

\newcommand\bcmdtab{\noindent\bgroup\tabcolsep=0pt%
  \begin{tabular}{@{}p{10pc}@{}p{20pc}@{}}}
\newcommand\ecmdtab{\end{tabular}\egroup}

\title[Parallel Logic Programming:  A Sequel]
{%
  Parallel Logic Programming: 
  A Sequel$^{\thanks{
      The authors would like to thank the anonymous reviewers for their
      careful reading and very valuable feedback. We would especially like
      to thank the editor, Mirek Truszczynski, for his very useful comments
      and encouragement.
      This research was partially supported by UNIUD PRID Encase, GNCS/INDAM
      grants, by NSF grants 1914635 and 1833630, by the Portuguese funding
      agency, FCT - Fundação para a Ciência e a Tecnologia, within project
      UIDB/50014/2020, by the Spanish MICINN project PID2019-108528RB-C21
      \emph{ProCode}, by the Madrid P2018/TCS-4339 \emph{BLOQUES-CM}
      program, and by the Tezos foundation.
    }}$
}

  \author[Dovier, Formisano, Gupta, Hermenegildo, Pontelli, Rocha]
         {AGOSTINO DOVIER \phantom{aaaaa}ANDREA FORMISANO\\
         	Universit\`a di Udine and GNCS-INdAM, Italy\\
         	\email{agostino.dovier|andrea.formisano@uniud.it}
	\and
         GOPAL GUPTA\\
         University of Texas at Dallas, USA\\
         \email{gupta@utdallas.edu}
         \and
         MANUEL V. HERMENEGILDO\\
         IMDEA Software Institute and Universidad Polit\'{e}cnica de Madrid, Spain\\
         \email{manuel.hermenegildo@\{imdea.org,upm.es\}}
         \and
                  ENRICO PONTELLI\\
         New Mexico State University, USA\\
         \email{epontell@cs.nmsu.edu}
         \and    
         RICARDO ROCHA\\
         CRACS/INESC TEC and Faculty of Sciences, 
         University of Porto, Portugal\\
         \email{ricroc@dcc.fc.up.pt}
         }

\jdate{July 2020}
\pubyear{2020}
\pagerange{\pageref{firstpage}--\pageref{lastpage}}
\doi{S1471068401001193}

\begin{document}

\label{firstpage}

\maketitle

  \begin{abstract}
 Multi-core and highly-connected architectures have become ubiquitous,
    and this has brought renewed interest in language-based approaches
    to the exploitation of parallelism.  Since its inception, logic
    programming has been recognized as a programming paradigm
    with great  potential for automated exploitation of parallelism.  The
    comprehensive survey of the first twenty years of research in parallel logic
    programming, published in 2001, has served since as a
    fundamental reference to researchers and developers.  The contents
    are quite valid today, but at the same time the field has
    continued evolving at a fast pace in the years that have followed.
    Many of these achievements and ongoing research have been driven by
    the rapid pace of technological innovation, that has led to
    advances such as very large clusters, the wide diffusion
    of multi-core processors, the game-changing role of 
    general-purpose graphic processing units, and the ubiquitous adoption of
    cloud computing. This has been paralleled by significant advances
    within logic programming, such as tabling, more powerful static
    analysis and verification, the rapid growth  of Answer Set
    Programming, and in general,  more mature implementations and systems.  This
    survey provides a review of the research in parallel logic
    programming covering the period  since 2001, thus providing a 
     natural continuation of the previous survey. In order to keep the 
     survey self-contained, it restricts its attention to parallelization of
     the major logic programming languages (Prolog, Datalog,  Answer Set Programming)
     and with an emphasis on automated parallelization and preservation of the
     sequential observable semantics of such languages. 
	The goal of the survey 
is to serve not only as a reference for researchers and developers of logic programming systems, but also as engaging reading for anyone interested in logic and as a useful source for researchers in parallel systems outside logic programming.
\\
\centerline{\em Under consideration in Theory and Practice of Logic Programming.}
  \end{abstract}
%
%
  \vspace*{4mm} 
  \begin{center}
    \begin{minipage}{.8\textwidth}
      \begin{quote}
        \emph{ In loving memory of our friends Francisco Bueno,
          Ricardo Lopes, and German Puebla, whose contributions to the
          field of logic programming have paved the way of many
          innovations and discoveries.  }
      \end{quote}
    \end{minipage}
  \end{center}
  \vspace*{4mm}

\section{Introduction}
\label{sec:intro}

The universal presence  of multi-core and highly-connected architectures 
has renewed interest in language-based approaches to the exploitation of parallelism.
Since its original inception, \emph{logic programming} has been recognized as
one of the programming paradigms with the greatest potential for automated
exploitation of parallelism. \citeANP{Kowalski79-LPS},  in his seminal book 
\cite{Kowalski79-LPS}
identifies parallelization as a strength of logic programming and
\citeANP{pollard-phd} explored the potential of 
the parallel execution of Prolog \cite{pollard-phd}. 
This was the beginning of an intense branch of research, with a number
of highlights  over the years---e.g., during the 
fifth generation computing systems project, 1982--1990.
A comprehensive survey of the first twenty years of parallel logic programming
has served for many years as a
reference to researchers and developers
\cite{parallellp-survey}. 
The content of that survey is still  valid today. At the same time,
the field of parallel and distributed logic programming has continued to evolve
at a fast pace,
touching on many exciting
achievements in more recent  years.

Many of these achievements and ongoing research have been driven by the
rapid pace of technological innovation, 
including the advent of Beowulf clusters, the wide diffusion of
multicore processors, the game-changing role of general-purpose Graphic Processing
Units (GPUs),  the wide adoption of cloud computing, and the advent of
big data with related computing infrastructures.

In the past 20 years, there has been a wealth of
additional research that has explored the role of new parallel architectures in
speeding up and scaling up the execution of different logic
programming paradigms.  Notable achievements have been accomplished in
the use of new generations of parallel architectures 
for both Prolog and 
Answer Set Programming (ASP), in the exploration of parallelism in constraint
programming, and in the extensive use of general purpose 
GPUs to speedup the execution of various 
flavors of logic programming.

This survey provides a review of the research in parallel logic
programming covering the period since 2000.
The survey has been designed to be self-contained and accessible, however,
being a natural continuation of the popular survey by \citeANP{parallellp-survey}, 
readers are strongly encouraged to read both surveys for a more comprehensive
perspective of the field of parallel logic programming.

The goal of the survey is to serve not only as a reference for
researchers and developers of logic programming systems, but also as
an engaging reading for anyone interested in logic
programming. Furthermore, we believe that this survey
 can provide useful insights and ideas to
researchers interested in parallel systems outside of the domain of
logic programming, as already happened in the past. We will
describe the key challenges encountered in realizing
different styles of parallelism in logic programming and review the most effective solutions
proposed in the literature. It is beyond the scope of this survey to review the actual
performance results produced---as they have been derived using a diversity of benchmarks, 
coding techniques, and hardware architectures. The interested reader will be able to find
such detailed results in the original papers cited in the bibliography of the survey.
We would also like to stress that our focus is primarily on systems
where parallelism does not modify the semantics of the programs being executed. The only exception is represented by the 
discussion on explicit parallelism, which introduces a different semantics with respect to the original Prolog systems.

A survey on the approaches to parallelism for constraint programming 
has been recently published \cite{DBLP:journals/tplp/GentMNMPMU18}. 
\emph{Constraint Logic Programming (CLP)} languages, if used for pure constraint modeling, 
can immediately benefit from the results presented in the constraint programming survey.
Considering that no recent work has appeared in the domain of parallelization of CLP, 
we will not discuss this area in this survey. 

Many of the systems that appeared in the literature and are mentioned in this survey
are publicly available. Interested readers are referred to the web page
\url{www.logicprogramming.org} (Systems and Links tab),
hosted by the Association for Logic Programming (ALP).

\medskip

The survey is organized as follows. Section~\ref{sec:background}
provides some background on logic programming and parallelism. A quick
review of the first 20 years of parallel logic programming is
presented in Section~\ref{sec:review}.  Section~\ref{sec:parprolog} 
explores the more recent advances in parallel execution of
Prolog, reviewing progress in execution
models for Or-parallelism (Section~\ref{sec:orpar}) and
And-parallelism (Section~\ref{sec:andpar}), static analysis for
exploitation of parallelism (Section~\ref{sec:staticanal}), and
finally exploitation of parallelism in Prolog systems with tabling
(Section~\ref{sec:tabling}).  Section~\ref{sec:asp} reviews the
techniques proposed to exploit parallelism in ASP, including
parallelism in Datalog (Section~\ref{sec:datalog}), traditional
parallelism in ASP (Section~\ref{sec:asppar}), parallel grounding
(Section~\ref{sec:parground}), and other forms of parallelism used in
ASP (Sections~\ref{otherASP}-\ref{sec:portfolio}). The following
sections explore the execution of logic programming in the context of
big data frameworks (``going large,'' Section~\ref{sec:bigdata}) and
in the context of GPUs (``going small,'' Section~\ref{sec:gpus}).
Section~\ref{sec:bigdata} also includes a discussion of execution
models for logic programming
on distributed computing platforms and frameworks designed
for handling massive quantities of data (e.g., MapReduce).
Section~\ref{sec:conclusion} provides some final remarks.

\section{Background}
\label{sec:background}

We start this section with a brief introduction to the foundations of logic programming.
Basically, a logic  program consists of facts and logical  rules, where
deduction is carried out by automatizing some variants of the \emph{modus ponens}.  
The idea is to  automatically derive the set of logical consequences of a logic program. 
If this set is finite, as is the case with the common restriction used in  ASP, a bottom-up
computation 
approach can be used.  
If, instead, this set is infinite, top-down computation methods allow us to 
derive selected inferences (e.g., for a given predicate). 
The interested reader is referred to the book by \citeN{lloyd} for a review of the fundamental
notions of logic programming. 
Some basic notions about parallelism and its limits are reported in Section \ref{sec:parall}.

\subsection{Logic Programming}
\label{sub:prel:lp}

A logic programming language is built on a signature composed of a set of function 
symbols $\cal F$, a set of variables $\cal X$, and a set of predicate symbols $\cal P$. 
An arity function $ar(p)$ is associated to each function
symbol and predicate symbol $p$.
A \emph{term} is either a 
variable $x \in {\cal X}$ or a formula of the form $f(t_1,\dots, t_n)$, where $t_1,\dots,t_n$ are themselves
terms, $f\in {\cal F}$, and $ar(f) = n \geq 0$. A constant is
 a symbol $f \in {\cal F}$ such that $ar(f)=0$. 

An atomic formula (or simply \emph{atom}) is a 
formula of the form $p(t_1,\dots,t_n)$, where $p \in {\cal P}$, $t_1,\dots,t_n$
are terms, and $ar(p)=n$. A \emph{literal} is either an atom or an entity of the
form $\textit{not }A$, where $A$ is an atom.
A \emph{clause} is a formula of the form 
\[ H \leftarrow B_1, \dots, B_n, \textit{not } C_1, \dots, \textit{not } C_m\]
where $H, B_1, \dots, B_n, C_1, \dots, C_m$ are atoms. 
If $m=0$ (i.e., there are no negated literals), the clause is said to be \emph{definite},
if $m=n=0$, then the clause is said to be a \emph{fact}.

Given a clause $r$, we define $head(r)=H$, $pos(r) = \{B_1,\dots,B_n\}$ and $neg(r)=\{C_1,\dots, C_m\}$, while $B_1, \dots, B_n, \textit{not } C_1, \dots, \textit{not } C_m$ is called the body of $r$.
The symbol ``$\leftarrow$'' can be read as ``if'';  in the rest of this paper we
will use $\leftarrow$ and its programming language syntactic notation
{\tt :-} interchangeably.  Intuitively, if the body 
of the clause is true, then the head will  be true as a consequence.
Variables are universally quantified.  
For instance, the clause
$$grandparent(X,Y) \leftarrow parent(X,Z), parent(Z,Y)$$ can be read as:
``\emph{for every $X,Y,Z$, if $X$ is  parent of $Z$ and $Z$ is parent of $Y$ then $X$ is a grandparent of $Y$}''.
A program is a collection of clauses. 
A program is \emph{definite} if all of its clauses are definite.
Negated literals are here written at the end of the body, for simplicity, but they can appear interleaved with positive literals in   actual  programs.

Important information concerning the semantics of a program $P$ can be obtained by the analysis of
its \emph{dependency graph}  $\mathcal{G}(P)$~\cite{baral03,gelfond_book}.
$\mathcal{G}(P)$  is a graph where nodes correspond to atoms  in $P$; an edge $p \leftarrow q$ is added
 in $\mathcal{G}(P)$ if there is a clause in $P$ with $p$ as head and $q$ in the body. The edge is labeled differently if $q$ occurs as an atom or a negated atom in the body of the clause.

A term (atom, clause, program) is \emph{ground} if it contains no variables.
A substitution $\theta$ is a mapping from a set of variables to terms.
If $r$ is a term (literal, clause) and $\theta$ a substitution, $r\theta$ denotes the term (literal, clause)
obtained by replacing each variable $X$ in $r$ with the term $\theta(X)$. 
Given two terms/atoms $s,t$, we say that $s$ is subsumed by $t$ (or $s$ is an instance of $t$)
if there is a substitution $\theta$ such that $s = t\theta$.
Two terms/atoms $s$ and $t$ are said to be \emph{variants} if $s$ is subsumed by $t$ and $t$
is subsumed by $s$. In this case, $s$ and $t$ are identical modulo a variable renaming.
An instance $t\theta$ of $t$ is a \emph{ground instance} if $t\theta$ contains no variables. 
Given a clause $r$,  $ground(r)$  denotes the set of all ground instances of $r$; 
analogously, given a program $P$, 
 $ground(P) = \bigcup_{r \in P} ground(r)$.

Given two substitutions $\sigma$ and $\theta$, $\sigma$ is more general than $\theta$ if there is
a substitution $\gamma$ such that 
for every term $t$ we have that $(t \sigma) \gamma = t \theta$.
$\theta$ is  a \emph{unifier} of two terms/atoms $s$ and $t$ if $s\theta = t\theta$, namely the two terms/atoms
become syntactically equal after the  substitution is applied.  If two terms/atoms admit a unifier, they 
will also  admit a 
\emph{most general unifier (mgu)}---i.e., the most general substitution which is also a unifier;  the
mgu is unique modulo variable renaming.

\medskip

Let us denote with $B_P$ the set of all possible ground atoms that can be built with the 
function and predicate symbols occurring in a program $P$,  also known as the
Herbrand Base of $P$.
An \emph{interpretation} $I\subseteq B_P$ is a set of ground atoms---intuitively representing which atoms are true;
all atoms in $B_P\setminus I$ are assumed to be false. 

An interpretation $I$ is a \emph{model} of a ground clause $r$ if either $head(r)\in I$, or $pos(r)\not\subseteq  I$,
or $neg(r)\cap I \neq \emptyset$ (namely, if the head is true or if the body is false).
An interpretation $I$ is a {model} of a clause if it is a model of all its ground instances.
An interpretation $I$ is a model of a program $P$ if it is a model of each clause in $P$.

The semantics of definite programs, which is used as the foundation of Datalog and Prolog, is 
defined in terms of
the set of logical consequences, namely by the set of atoms that are true in every model.
For a definite logic program $P$ the existence of a unique minimum model, denoted by $M_P$, is guaranteed. 
$M_P$ can be defined as the intersection of all $I$ such that $I$ is a model of $P$; it can be shown that
$M_P$ corresponds to the set of all ground logical consequences of $P$.

The minimal model $M_P$ of a definite logic program has a constructive characterization through the
use of the \emph{immediate consequence operator} $T_P$. The operator maps interpretations to interpretations, 
and it is defined as:
$$T_P(I) = \{ head(r)  \:|\:  \mbox{$r$ ground instance of a clause in $P$},\, pos(r) \subseteq I\}$$
$T_P$ is a monotone and continuous operator; in addition,
$M_P$ is the least fixpoint of $T_P$, i.e., $M_P = T_P(M_P)$, and $M_P = T_P \uparrow\omega$, where
$T_P\uparrow 0=\emptyset$, $T_P \uparrow n = T_P (T_P \uparrow (n-1))$ for a successor ordinal $n$, 
and $T_P \uparrow \alpha = lub \{T_P\uparrow n \:|\: n < \alpha\}$ for a limit ordinal $\alpha$. Computing $M_P$ as $T_P \uparrow\omega = 
T_P(T_P( \cdots (T_P(\emptyset))\cdots ))$ is called the \emph{bottom-up approach} to the semantics of $P$.

\smallskip

In the context of Prolog, computation is expressed in terms of reasoning about the  minimal
model $M_P$. Given a program $P$ and a conjunction of possibly non-ground atoms $\bigwedge_{i=1}^n p_i$, the objective is
to determine substitutions $\theta$ for the variables in $p_1, \dots,p_n$ such that 
$M_P \models \bigwedge_{i=1}^n p_i\theta$; these are referred to as \emph{correct answers}. The conjunction
$\bigwedge_{i=1}^n p_i$ is referred to as a \emph{goal} and subsets of $\{p_1,\dots,p_n\}$
 are called \emph{subgoals} (let us observe that a subgoal can contain more than one atom).

SLD resolution is a strategy used to derive correct answers. It can
be described as the process of
constructing a tree, the SLD-resolution tree, whose nodes are pairs composed of a goal and a substitution.  Given a 
goal $\vec{p}=\bigwedge_{i=1}^n p_i$ and a program $P$, the root of the resolution tree is $\langle \vec{p}, \epsilon\rangle$, where
$\epsilon$ is the identity substitution. If 
$\langle \bigwedge_{i=1}^m q_i, \theta \rangle$ is a node in the resolution tree, then such node
will  have as many children as there are rules $r\in P$
 such that there is a substitution $\gamma$ such that $head(r')\gamma = q_1\theta\gamma$, where 
  $r'$ is a renaming of the rule $r$ with fresh new variables. 
  In this case a most general unifier of $head(r')$ and $q_1\theta$ is chosen.
 Each  child is of the form
 $\langle pos(r') \circ [q_2,\dots,q_n], \theta\gamma\rangle$, where  $\theta\gamma$ is the composition of the
  two substitutions and $\circ$ is the  list concatenation operator.
 If $m=0$, then the node will be labeled as a \emph{success node} and $\theta$ is a correct answer.
 If $m>0$ but no child  can be constructed, then the node will be labeled as
 a \emph{failed node}.  Prolog typically
 implements SLD resolution by building the resolution tree in a depth-first manner, where the
 children of each node are sorted according to the ordering of the clauses in the program $P$. 
The SLD resolution provides a \emph{top-down} approach
to reason about  the semantics of $P$. 

The top-down approach is extended in the case of negated literals in goals (general programs).
The goal $\textit{not } p$ succeeds or fails as a consequence of the result of the goal $p$. 
If  there is an answer substitution $\theta$ for $p$  then the goal $\textit{not } p$ fails,  
otherwise it succeeds. This is referred to as \emph{negation as failure}.

However, for programs with general clauses, the existence and 
uniqueness of a minimum model is no longer guaranteed.
The  \emph{well-founded} model of a logic program  is a unique 3-valued model \cite{wfs} or 4-valued model \cite{DBLP:books/mc/18/Truszczynski18}.
We briefly report the procedure presented by 
\citeN{DBLP:journals/tplp/BrassDFZ01},
based on the following variant of the $T_P$ operator, for computing the 3-valued well-founded semantics.
Given a ground program $P$ and two sets of ground atoms $I,J$, we define the extended immediate consequence operator
$$T_{P,J}(I) = \{head(r) \:|\: r \in P, pos(r) \subseteq I,neg(r)\cap J=\emptyset\}$$ 
Let us denote by $P^+$ the set of definite clauses in $P$, the \emph{alternating} fixpoint procedure is defined as follows
\[
\begin{array}{lcl}
	K(0) = \mathit{lfp}(T_{P^+}) &  & U(0) = \mathit{lfp}(T_{P,K(0)})\\
	K(i+1) = \mathit{lfp}(T_{P,U(i)}) && U(i+1) = \mathit{lfp}(T_{P,K(i+1)})
\end{array}
\]
Let $(K^*,U^*)$ be the minimum fixpoint of the computation; that is,  if $i$ is the smallest value such that 
$K(i)=K(i-1)$ and $U(i)=U(i-1)$, then 
$K^*=K(i)$ and $U^*=U(i)$.

The well-founded model of the program $P$ is a 3-valued model  $W^*$ defined as
$W^* = (K^*, U^*)$---where all atoms in $K^*$ are true, all atoms in $U^*$ are false,
and all atoms in $H \setminus (K^*\cup U^*)$ are unknown.

The most successful approach for the two-valued semantics
is the one based on the  notion of  \emph{answer sets}. 
Given a program $P$ (for simplicity let us assume $P$ to be ground) and given an interpretation $I$, we define the
reduct of $P$ with respect to $I$, denoted by $P^I$ as the set of definite clauses 
$P^I = \{head(r)\leftarrow pos(r)\:|\: r\in P, neg(r)\cap I=\emptyset\}$. 
A model $M$ of $P$ is an \emph{answer set} of~$P$ if $M$ is the minimum model of the definite program $P^M$. 
It should be noted that a program may have no answer sets, one answer set (for example, each definite program has 
exactly one answer set, corresponding to its minimal model), or multiple answer sets.

\subsection{Parallelism and Speedups}\label{sec:parall}

When talking about parallelism, 
it is a naive  belief that if a program runs in time $T$ on a computing platform, then   
 it should be possible to execute the program  $N$ times faster using $N$ processors.\footnote{In the rest of the paper, we will use the terms \emph{processor} and \emph{process} in a general sense, as representing an entity capable of computation (e.g., a CPU, a core).} 
Namely, that the running time could decrease to $\frac{T}{N}$.

A definitive theoretical limit to this kind of reasoning was set by \citeN{DBLP:conf/afips/Amdahl67}. 
The crucial point is that the program we are considering is
composed of parts that are intrinsically sequential and  other parts that can be parallelized.  Let $S$ (sequential) and $P$ (parallel) be the running times of the two fractions of the program, scaled to guarantee that  $S+P=1$.
With $N$ processors one can expect an ideal  running time of $T\left(S + \frac{P}{N}\right)$,
with a speedup of $\frac{1}{S + \frac{P}{N}}$.
Observe that with a ``very large'' $N$ this leads to a maximum speedup 
of $\frac{1}{S}$. Even if $S$ is ``small'' with respect to the program, e.g., $S=0.1$, this means that the maximum speedup is 10, no matter how many processors are used.
The naive reasoning mentioned at the beginning of the paragraph assumes implicitly that $S=0$, which is 
often not a realistic assumption.

\citeN{DBLP:journals/cacm/Gustafson88}
approached the problem from another perspective, giving new hopes of high impact for parallel systems.
The key observation of Gustafson is that
Amdahl's law assumes that the problem size (the data size) is
fixed. However, Gustafson argues that, in practice, it makes more
sense to scale the size of the problem (e.g., the amount of data) in
parallel with the addition of processors.
Given the running time $T$ as discussed earlier, let 
 us assume that the parallel part of the program $P$ has ``to
do with data,'' and that the architecture allows
parallel processors to concurrently access 
independent portions of
the data. Then in the same time $T$, in principle, the program could
process $N$ times more data, leading to a speedup
of $\frac{PN+S}{P+S}$.
If $S=0.1$, with one thousand of processors we can, in principle, aim at  
reaching a speedup of $\approx 900$.


\section{The First 20 Years of Parallel Logic Programming: A Quick Review}
\label{sec:review}


The majority of the original research conducted on parallel execution
of logic programming focused on the exploitation of parallelism from the execution of
Prolog programs.
In this section, we provide a brief summary of some
of the core research directions explored in the original literature.
Readers are  encouraged  to review the survey by \citeN{parallellp-survey} 
for further details.

\subsection{Explicit Parallelism}


Parallel execution of the different operations in a logic program
results in {\it implicit} exploitation of parallelism, i.e., no
input is required from the user to identify and exploit parallelism; rather,
parallelism is automatically exploited by the inference system.
Nevertheless, the literature has presented several  approaches that explore
 extensions of  a logic programming
language with \emph{explicit} constructs for the description of 
concurrency and parallelism. 
While these approaches are not the focus of this survey, we will briefly
mention some of them in this section.  

The  approaches for the  explicit description of parallelism 
in logic programming can be largely
classified into three categories: {\bf (1)} \emph{message passing}; 
{\bf (2)} \emph{shared memory}; and {\bf(3)} \emph{data-flow}.

Methods based on \emph{message passing} extend a logic programming language, typically
rooted in Prolog, to enable the creation of concurrent computations and the communication
among them through the explicit exchange of messages.
Several systems have been described over the years with similar 
capabilities, such as Delta Prolog \cite{pereira86} and CS-Prolog \cite{futo93}.
A more recent  example is the April system~\cite{Fonseca-06}, where the
high-level modeling capabilities of logic programming are used to
explicitly parallelize Machine Learning procedures into independent
tasks, in the various stages of data analysis, search, and evaluation
\cite{DBLP:journals/ml/FonsecaSSC09}.

Message passing features have  become
common in many implementations of Prolog---for example,
Ciao Prolog~\cite{ciao-design-tplp-short},
SICStus Prolog~\cite{sicstus-journal-2012-short},
SWI-Prolog~\cite{swi-journal-2012}, tuProlog~\cite{lpaas-tplp18},
XSB~\cite{Swift-12}, and YAP Prolog~\cite{yap-journal-2012}.

Most Prolog implementations provide comprehensive
multi-threading libraries, allowing the creation of separate threads
and enabling message passing communication between them so (see also
the paper by \citeN{PROLOGPASTPRESENTFUTURE2021}). The design of these libraries has
also been guided by a 2007 ISO technical document which provides
recommendations  on
how multi-threading predicates may be introduced in Prolog.

\noindent
E.g., in YAP Prolog threads are created as follows:

{
\begin{quote}
\tt	thread\_create(:Goal, -Id, +Options)
\end{quote}
}

\noindent and the communication between them can be 
achieved  through message queues, e.g.,

{
\begin{quote}
\tt	thread\_send\_message(+ThreadId, +Term)
	
	thread\_get\_message(?Term)
\end{quote}
}

The roots of message-passing Prolog systems date back to the early 80s. The Delta Prolog system~\cite{PeNasr85} draws inspiration from communicating sequential processes to provide the ability
of creating processes capable of synchronizing and exchanging messages; for example, the following
code snippet describes a simple producer-consumer structure:

{
\begin{quote}
\tt  main :- producer // consumer. 

  producer :- generate(X), X ! channel, producer. 
  
  consumer :- X ? channel, consume(X), consumer.
\end{quote}
}
A more extensive design is proposed in the CS-Prolog system \cite{csprolog}, which provides communicating process designed to execute on Transputer architectures. 

\medskip
The second class of methods is based on the use of \emph{shared memory} 
to enable communication between concurrent processes. The most notable approaches rely on 
the use of blackboards, as exemplified in the Shared Prolog system \cite{ciancarini} and in the
various Linda libraries available in several Prolog implementations (e.g., SICStus). 
An interesting alternative is provided by Ciao Prolog, where the communication
is done using \emph{concurrent facts} in the Prolog
database~\cite{shared-database}. This approach is designed to provide a clear transactional 
behavior, elegantly interacting with  backtracking and preserving, under well-defined
conditions, the declarative semantics of logic programming.

\medskip
Finally, a more extensive literature has been developed around the concept of \emph{concurrent logic programming} (see for example the works by
\citeANP{cp-collect}~\citeyear{cp-collect,Shapiro-survey},\citeN{Parlog-Acm}, and \citeN{tick-deev}) and implemented in systems like Parlog and GHC. These languages
offer a syntax similar to traditional logic programming, but they view
all goals in the program clauses as concurrent processes, capable of interacting through shared variables and 
synchronizing through
dataflow constraints. For example, a goal of the type ~{\tt ?- generate(Z), proc(Z)}~ creates two concurrent processes,
 the first generating values in a list (used as a stream in concurrent logic programming) and the 
 second process  consuming values from the stream, e.g., 
{
\begin{quote}
\tt  proc([X|Y]) :- X > 0  | consume(X), proc(Y).

  proc([X|Y]) :- X <= 0 | proc(Y).
\end{quote}
}
The second process suspends until a value is available in the stream; the clause whose guard (i.e., the 
goal preceding the ``$|$'') is satisfied  will be activated and executed.
One important aspect to observe is that, due to the complications of
combining search with this type of concurrency, these systems do not support
the classical non-determinism of logic programs. They instead implement
only the deterministic part---a form of computation referred to as
``committed choice.''
The fact that all goals are concurrent processes may also lead to  
complex and unintended interactions, making programs harder to understand.

The idea of synchronization through shared variables is very
appealing. This concept has generated several 
research efforts focused on  supporting
variable binding-based communication in conjunction with explicit
concurrency in Prolog systems, e.g., in several proposals for 
 dependent
And-parallelism \cite{pasco,att-var-iclp-short,kish-jlp}. More 
information about these directions of work can be found later in 
this section and in  Section~\ref{sec:andpar}.


\subsection{Implicit Parallelism}
The fundamental idea underlying the implicit exploitation of parallelism is to make the exploitation
of parallelism transparent to the programmer, by automatically performing in parallel some of the
steps belonging to the operational semantics of the language.

As mentioned earlier, most of the pre-2000 literature on implicit exploitation of
parallelism focused on the parallelization of the execution model
of Prolog. In very general terms, the traditional SLD-resolution process adopted by Prolog can be
visualized as a search procedure (see  Algorithm~\ref{alg:sld}). 
There are three major steps in Algorithm 1: \emph{Atom Selection,} \emph{Clause Selection,} and 
\emph{Unification.} 
Atom Selection (also known as literal selection in literature)
is used to identify the next atom to resolve (line~3 of Algorithm~\ref{alg:sld}). 
Clause Selection is used to identify which clause to use to conduct resolution (line~5). Unification
determines the most general unifier to be used to unify the selected literal with the head of the
selected clause (line~6).

Extensive research has been performed to improve efficiency of  these steps, especially,
for atom selection and clause selection. Techniques such as constraint programming
\cite{DBLP:journals/tplp/GentMNMPMU18} and the Andorra principle \cite{andorra-I-engine}
attempt to select subgoals in an order that will lead to early pruning of the search space. 
The parallelization of the atom selection step leads to 
\emph{independent} and \emph{dependent And-parallelism} \cite{Hampaper,kish-jlp,pasco}, which allow
 multiple subgoals to be selected and solved in parallel. The distinction between the two forms of 
 And-parallelism 
 derives from whether shared unbound variables are allowed (dependent And-parallelism) or not 
 (independent And-parallelism) between concurrently 
 resolved subgoals. 
The parallelization of the Clause Selection step,
which allows resolution of the selected subgoal using multiple clauses in parallel, leads to 
\emph{Or-parallelism} \cite{zhang93,aurora,muse}.

\begin{algorithm}[htbp]
\DontPrintSemicolon
\KwInput{Input Goal}
\KwInput{Program Clauses}
\KwOutput{Computed Substitution}

$i=0$

\While{(Goal not Empty)}
{
	{\it select}$_\textit{atom}$ B from Goal  \tcc*{And-Parallelism}
	\Repeat{(unify(H,B) {\bf or} no clause is left)\tcc*[f]{Unification Parallelism}}
	{
		{\it select$_\textit{clause}$} (H :- Body) from Program \tcc*{Or-Parallelism}
	} 
	\If{(no clause is left)}
	{
		\Return {\bf Fail}\;
	}
	\Else
	{
		$\theta_i$ = most\_general\_unifier(H,B)\; 		
		Goal = (Goal $\setminus$ \{B\} $\cup$ Body)$\theta_i$\;
		$i$++\;
	}
}
\Return  $\theta_0\theta_1\cdots \theta_{i-1}$\;

\caption{High-level view of SLD-Resolution}
\label{alg:sld}
\end{algorithm}

{\it Unification parallelism} arises when arguments of 
a goal are concurrently unified
with those of a clause head with the same name and arity. This can be
visualized in Algorithm~\ref{alg:sld} as the parallel execution of the
steps present in the \emph{unify} operation.
The different
argument terms can be unified in parallel as can the different subterms 
in a term \cite{barklund-phd}. 
Unification parallelism is very fine-grained, it requires dealing with dependencies
caused by multiple occurrences of the same variables,  and is
best exploited by building specialized processors with multiple
unification units \cite{parall-unif}. 
Unification parallelism has not been the major focus of 
research in parallel logic programming, so we will not consider it any further.

\subsubsection{Or-Parallelism}
{\it Or-parallelism} arises when multiple clauses have the same predicate
in the head and the selected subgoal
unifies with more than one clause head---the corresponding clause bodies
can  be explored in parallel, giving rise to Or-parallelism.  This can
be illustrated in Algorithm~\ref{alg:sld} as the parallelization of
the choices present in the \emph{select}$_\textit{clause}$ operation.
Or-parallelism is thus a way of efficiently searching
for solutions to the goal, by exploring alternative
solutions in parallel: it corresponds
to the parallel exploration of the search tree originating from the
choices performed when selecting different clauses for solving the
selected subgoals (often referred to as the \emph{or-tree}).

Let us consider a very simple example, composed of the Prolog clauses
that define the concatenation of two lists:

{
\begin{quote}
\tt append([], X, X).
	
	append([H|X], Y, [H|Z]) :- append(X, Y, Z).
\end{quote}
}

\noindent and the goal:
{
\begin{quote}
\tt ?- append(X, Y, [1,2,3]).
\end{quote}
}

\noindent
The {\tt append} literal in the goal (and all {\tt append} subgoals
that arise recursively during resolution) will match both clauses, and
therefore the bodies of these clauses can be executed in parallel to
find the four solutions to this goal.%
\footnote{Note however that in this example exploiting parallelism may
  not be too profitable since the \emph{granularity} of the tasks is
  small. We will return to this important topic.}

\begin{figure}[htbp]
	\begin{center}
\begin{tikzpicture}[xscale=0.5,yscale=0.5]\small
\thicklines
\draw  (5,15) ellipse (6 and 1);
\draw  (5,15) node  {{\tt ?- append(X, Y, [1,2,3]).}};
\draw [dashed, line width=1mm, gray, ->] (4,14) -- (2,11);
\draw [dashed, line width=1mm, gray, ->] (6,14) -- (8,11);
\draw  (-0.3,12.5) node  {{\tt append([], X1, X1).}};
\draw  (13,12.5) node  {{\tt append([H2|X2], Y2, [H2|Z2]) :-}};
\draw  (15,11.8) node  {{\tt  append(X2, Y2, Z2).}};

\draw (0,10)   ellipse (4 and 1);
\draw (10,10)  ellipse (5 and 1);

\draw  (-.5,10) node  {{\tt X=[], Y= [1,2,3]}};
\draw  (10,10) node  {{\tt ?- append(X2, Y2, [2,3]).}};

\draw [dashed, line width=1mm, gray, ->] (6,9.3) -- (5,7);
\draw [dashed, line width=1mm, gray, ->] (13,9.1) -- (14,7);
\end{tikzpicture}

	\caption{Example of Or-parallelism}
	\label{fig:orp}
	\end{center}
\end{figure}
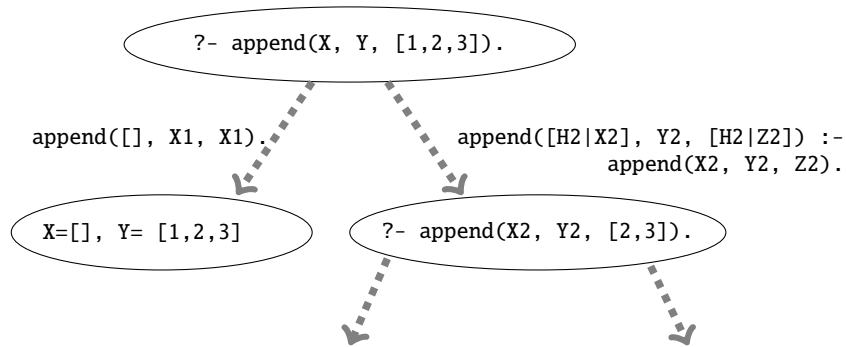

Or-parallelism should be, in principle, easy to achieve, since the
various branches of the or-tree are independent of each other, as they
each explore an alternative sequence of resolution steps. As such,
their construction should require little communication between
parallel computations.\footnote{Minor control synchronization is still
  needed. Moreover, we are not considering the challenge of dealing
  with side-effects, which require communication between branches of
  the or-tree.}  However, in practice, implementation of
or-parallelism is difficult because of the sharing of nodes in the
or-tree. Given two nodes in two different branches of the or-tree, all
nodes above (and including) their least common ancestor node are
shared between the two branches. A variable created in one of these
ancestor nodes might be bound differently in the two branches. The
environments of the two branches have to be organized in such a way
that, in spite of the ancestor nodes being shared, the correct
bindings applicable to each of the two branches are easily
discernible.

If a binding for a variable, created in one of the common ancestor 
nodes, is generated above (or at) the least common ancestor node, then
this binding will be the same for both branches, and hence can be used
as such. Such binding is known as an \emph{unconditional binding} and
such a variable is referred to as an unconditional variable.
However, if a binding to such a variable 
is generated by a node below the least 
common ancestor node, then that binding shall be visible
only in the branch to which the binding node belongs. Such
a binding is known as a \emph{conditional binding} and such a variable
is referred to as 
a conditional variable. The main problem in implementing
Or-parallelism is the efficient representation of multiple
environments that co-exist simultaneously in the or-tree---commonly
known as the \emph{environment representation problem}. 
Note that the main problem
in the management of multiple environments is how to efficiently 
represent and access the conditional bindings;
the unconditional bindings can be treated as in normal sequential
execution of logic programs. The environment representation
problem  has
to be solved by devising a mechanism where each branch has some
private areas where it stores the conditional bindings applicable
to such branch. Several approaches have been 
explored to address this problem. For example:

\begin{description}
	
	\item[$\bullet$] By storing the conditional bindings created
	by a branch in an array or
	a hash table private to that branch,
	from where the bindings are accessed whenever they are needed. This 
	approach has been adopted, for example, in the binding array 
	model, successfully used in the Aurora system \cite{Warren84,aurora}.
	
	\item[$\bullet$] Keeping a separate copy 
	of the environment for each branch of the tree, so that 
	every time branching
	occurs at a node the environment of the old branch is
	copied to each new branch. This approach has been adopted, for
	example, in the stack-copying model, successfully used in the
	Muse and the ACE systems \cite{Ali-90a,annals}.
		
	\item[$\bullet$] Recording all the conditional
	bindings in a global data-structure and attaching
	a unique identifier with each binding
	which identifies the branch a binding belongs to. This
	approach has been explored, for example, in the version
	vectors model \cite{versionvectors}.
	
\end{description}

\subsubsection{And-Parallelism}
{\it And-parallelism} arises when more than one subgoal is present
in the goal or in the body of a clause, and multiple subgoals
are selected and resolved concurrently. This can be visualized in 
Algorithm~\ref{alg:sld} as generating parallelism from the operation
\emph{select}$_\textit{atom}$.
The literature has traditionally distinguished between \emph{independent
	And-parallelism} and \emph{dependent And-parallelism}.

\medskip
In the case of \emph{independent And-parallelism}, subgoals selected for parallel
execution are guaranteed to be independent of each other with respect to
bindings of variables. In other words, two subgoals solved in parallel are guaranteed
to not compete in the binding of unbound variables.  
A sufficient condition for this (called \emph{strict independence}) is that
any two subgoals solved in parallel do not have any unbound variables in 
common.
Independent And-parallelism is, for example, a way of speeding up a divide-and-conquer
algorithm by executing the independent subproblems in parallel. The advantage
of independence is that the parallel execution can be carried out without
interaction through shared variables. Communication is still needed for
returning values and synchronization (including during backtracking).

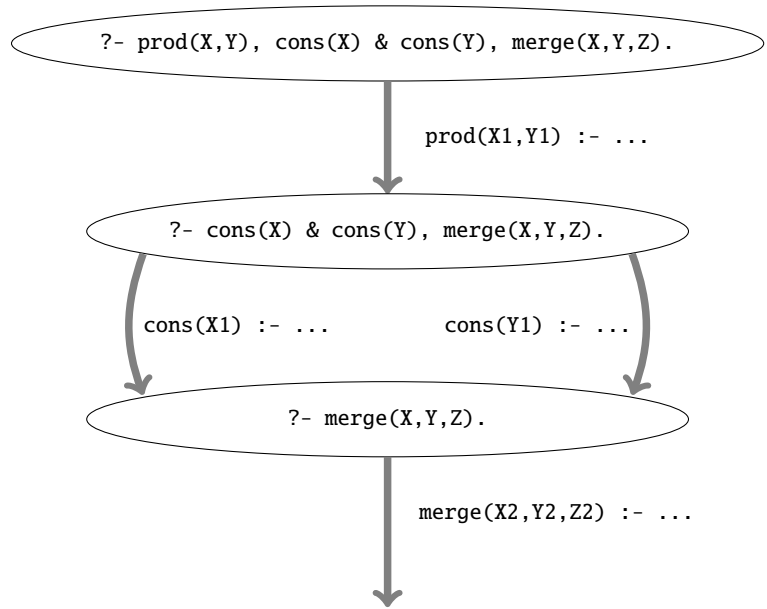
\begin{figure}[htbp]
	\begin{center}
\begin{tikzpicture}[xscale=0.5,yscale=0.5]
\thicklines \small
\draw   (5,15) ellipse (10 and 1);
\draw  (5,15) node  {{\tt ?- prod(X,Y),  cons(X) \& cons(Y), merge(X,Y,Z).}};
\draw [line width=1mm, gray, ->] (5,14) -- (5,11);
\draw  (9,12.5) node  {{\tt prod(X1,Y1) :- ...}};

\draw (5,10)    ellipse (8 and 1);
\draw  (5,10) node  {{\tt ?- cons(X) \& cons(Y), merge(X,Y,Z).}};

\draw  (1,7.5) node  {{\tt cons(X1) :- ...}};
\draw  (9,7.5) node  {{\tt cons(Y1) :- ...}};

\draw [line width=1mm, gray,-> ] (-1.5,9.4) .. controls (-2,8) and (-2,7) .. (-1.5,5.65);
\draw [line width=1mm, gray, ->] (11.5,9.4) .. controls (12,8) and (12,7) .. (11.5,5.65);
\draw     (5,5) ellipse (8 and 1);
\draw  (5,5) node  {{\tt ?- merge(X,Y,Z).}};
\draw  (9.5,2.5) node  {{\tt merge(X2,Y2,Z2) :- ...}};
\draw [line width=1mm, gray,->] (5,4) -- (5,0);
\end{tikzpicture}
	\caption{Intuition behind independent And-parallelism}
	\label{fig:andp}
	\end{center}
\end{figure}

The main systems proposed to support independent And-parallelism,
e.g., \&-Prolog \cite{Hampaper} and \&ACE~\cite{santa-barbara},
adopted initially a fork-join organization of the computation,
inspired by the original proposal on \emph{restricted} And-parallelism
by \citeN{DeGroot84}. In this fork-join model
independent goals selected to be run in parallel
(identified, e.g., using a different conjunction operator \&) are made 
available simultaneously for parallel execution, and the
continuation of the computation waits for completion of these 
parallel goals before proceeding
(e.g., see Figure \ref{fig:andp}).  
These And-parallel structures can be arbitrarily nested. 
%

\medskip
The major challenges in the development of independent And-parallel systems are:
\begin{itemize}
	\item 
The
implementation of distributed backtracking---the goal of maintaining the same visible 
behavior as a sequential Prolog system implies the need of allowing a processor
to backtrack over computations performed by other processors, with requirements of communication
and synchronization; these issues have been extensively explored by
\citeN{hermenegildo-phd-short}\citeN{Backtracking}\citeN{flexmem-europar96}\citeN{ace-backtrack-tpds};
\citeANP{parback-chico-ICLP11}\citeyear{parback-chico-ICLP11,chico-casas-padl12}.

\item The identification of subgoals that will be independent at run-time---this problem has
  been addressed by Hermenegildo~\citeyear{hermenegildo-phd-short,Hampaper} 
  through manual program annotations 
  and by  \citeN{annotators-jlp} using static analysis techniques.

\item The need to allow more relaxed notions of
  independence, i.e.,  the conditions that  guarantee that subgoals can be solved
  in parallel without communication. These include:
  \begin{itemize}
  	\item the classical notion of
  	independence (called \emph{strict
          independence} by~\citeN{sinsi-jlp},
        i.e., goals should not share variables at run time;
  	\item \emph{non-strict independence}, characterized by the fact that
          only one goal can bind each shared
  	variable~\cite{sinsi-jlp};
  	\item \emph{search independence}, i.e., bindings are
  	compatible~\cite{maria-phd};  
  	\item \emph{constraint
  	independence}~\cite{consind-toplas}.
  \end{itemize} 
Deterministic behavior is also a form of independence~\cite{sinsi-jlp}.
\end{itemize}

\bigskip	
\noindent {\it Dependent And-parallelism} arises when two or more subgoals
that are executed in parallel have variables in common 
and the bindings made by one subgoal affect the execution of  other subgoals.  Dependent And-parallelism can be
exploited in two ways (let us consider the simple case of two subgoals):
\begin{enumerate}
	\item\label{cone} The two subgoals can be executed independently until
one of them accesses/binds the common variable. 
It is  possible to continue executing 
the two subgoals independently 
in parallel, separately maintaining the bindings generated by each subgoal.
In such a case, at the end of the execution of the two subgoals,
 the bindings
produced by each will have to be checked for compatibility (this
compatibility check at the end is referred to as {\it back unification}).
\item\label{ctwo}  Once the common variable is accessed by one of the subgoals, it is
bound to a structure, or {\it stream} (the goal generating this
binding is called the {\it producer}), and the structure is
read as an input argument by the other goal (called the {\it consumer}).
\end{enumerate}
Case {\bf(\ref{cone})} is very similar to independent And-parallelism
and can be seen as exploiting independence at a different granularity
level. 
Case {\bf(\ref{ctwo})}
is sometimes also referred to as {\it stream-parallelism} and is
useful for speeding up producer-consumer
interactions, by allowing the consumer
goal to compute with one element of the stream while the producer
goal is computing the next element. Note
that stream-parallelism introduces a form of coroutining. 
Stream-parallelism forms the basis of the Committed Choice 
Languages mentioned earlier (e.g., Parlog \cite{Parlog-Acm}, GHC \cite{ueda-phd}, and 
Concurrent Prolog (Shapiro \citeyearNP{cp-collect,Shapiro-survey})). 

\medskip
The main challenge in implementing dependent And-parallelism is 
controlling the parallel execution of the consumer subgoal. Given
 two subgoals that share a variable {\tt X}, one a 
producer of {\tt X}
and another a consumer, the execution of both can be
initiated in parallel. However, one must make sure that the
consumer subgoal computes only as far as it does not instantiate
the dependent variable {\tt X} to a non-variable binding. 
If it attempts to do so,
it must suspend. It will be woken up only after a non-variable 
binding has been produced for {\tt X} by the producer subgoal---or,
alternatively, after the producer has completed its execution without
binding {\tt X}. 
Thus, in addition to the concerns mentioned in the context of
independent And-parallelism, the two additional main concerns in
implementing dependent And-parallelism are: 

\begin{enumerate} 
	
	\item Determining
	which instance of a given dependent variable is a producer
	instance and which instances are consumer instances; 
	
	\item Developing efficient
	mechanisms for waking up suspended subgoals when the variables
	on which they were suspended are instantiated (or turned themselves
	into producer instances). 
	
\end{enumerate} 
In Prolog execution, subgoals in the current goal 
are resolved in a left-to-right order.
Thus, when a subgoal is
resolved, it will never have an unexecuted subgoal to its left
in the goal.  This rule, however,
can be relaxed, and considerable advantage can be gained by processing
goals in a different order.
In particular, this can be applied to subgoals that
have at most one matching clause, known as {\it determinate}  subgoals.
This has been realized in the \emph{Andorra principle} \cite{andorra-I}. 
The Andorra principle states that
all determinate subgoals in the current
goal should be executed first, irrespective of their position in the goal.
Once all determinate subgoals have finished execution, the leftmost
non-determinate subgoal is selected and its various alternatives
tried in the standard Prolog order. Along each alternative, the same principle
is applied. 
Under the Andorra principle, all deterministic decisions are taken as soon as possible and this 
facilitates coroutining and leads to a significant narrowing of the search space.

The Andorra principle has been realized in the Andorra-I system \cite{andorra-I-engine}.
The Andorra-I system also has a determinacy analyzer which generates conditions
for each subgoal at compile-time \cite{andorra-I-prep}. 
These simple conditions are checked at
runtime and their success indicates that the corresponding subgoal is determinate.
The Andorra-I system is a {\it goal stacking} implementation of Prolog rather than the traditional
{\it environment stacking}, due to the need to reorder subgoals
during execution.
The Andorra principle is in fact useful even beyond parallelism, as a
control rule for Prolog, and can also be implemented using delay
primitives~\cite{qe-andorra-fim}. This is supported for example by the
Ciao Prolog system.

The Andorra principle has been generalized to the Extended Andorra
Model (EAM) \cite{HaridiBrand88} to exploit both And- and Or-parallelism.
In EAM, arbitrary subgoals can execute in And-parallel and clauses can
be tried in Or-parallel. 
Computations
that do not impact the environment external to a subgoal are freely executed,
giving rise to parallelism. However, computations that may bind a variable
occurring in argument terms of a subgoal are suspended, unless the binding
is deterministic (in which case the binding is said to be \emph{promoted}).
If the binding is non-deterministic, then the subgoal is replicated for each
binding (\emph{non-determinate promotion}). These replicated subgoals are
executed in Or-parallel. At any moment, constraints and bindings
generated outside of a subgoal are immediately percolated down to that
subgoal (propagation).

The Extended Andorra Model seeks to optimally exploit And-/Or-parallelism.  
It provides a generic model for the exploitation of
coroutining and parallelism in logic programming and has motivated two main lines
of research. The first path resulted in the Andorra Kernel Language
\cite{akl} that can be thought of as a new paradigm that subsumes both
Prolog and concurrent logic languages \cite{HaridiBrand88}.
The second focused on the
EAM with Implicit Control \cite{EAM,BEAMTPLP2011}, where the goal is
to achieve efficient (parallel) execution of logic programs
with minimal programmer control. An interpreter (in Prolog) was developed to
better understand EAM with implicit control \cite{eam-gupta-warren} and
new concepts, such as lazy copying and eager producers, that give finer
control over search and improve parallelism have been investigated. Gupta and Pontelli
later experimented with an extension of dependent And-parallelism that
provides some of the functionality of the EAM through parallelism
\cite{pasco}. The first prototype parallel implementation
of the EAM (called BEAM), based on the WAM, has been presented by  \citeANP{beam1}\citeyear{beam1,beam2}.
The BEAM system is  promising, though
the fine grain And-/Or-parallelism that EAM supports results in significant
overhead and thus extracting good performance requires taking into account
additional factors such as granularity control.

The state-of-the-art in the exploitation of dependent And-parallelism
at the beginning of 2000 is represented by systems like
ACE~\cite{pasco}, BEAM~\cite{BEAMTPLP2011} and DDAS~\cite{kish-jlp}, which are effective but
complex.  There are also some early attempts at simpler approaches to parallel implementations~\cite{ciao-ilps95}. 


\section{Parallel Execution of Prolog} \label{sec:parprolog}

Following the brief review made in the previous section of some of
the core issues underlying parallel execution of Prolog, 
let us now turn our attention to the  advances made in the area of parallel execution 
of logic programming since 2000. We start, in this section, 
with reviewing the most recent progress in the context of parallel execution of Prolog.
We survey  new theoretical and practical results in Or- and
And-parallelism, discuss advances in Static Analysis to aid 
with the exploitation of parallelism, and, finally, discuss
the combination of parallelism with  the notion of Tabling.

\subsection{Or-Parallelism}
\label{sec:orpar}


\subsubsection{Theoretical Results}

The literature on
Or-parallel execution of Prolog is extensive and it is primarily focused
on the development of solutions for the
\emph{environment representation problem}. More than twenty different
approaches have been presented in the literature  to address this problem.

From the theoretical point of view, the environment 
representation problem has been formalized as a data structure problem over labeled trees \cite{op-ngc}.
The Or-parallel execution  of a program can be abstracted as building a labeled
tree,\footnote{Without loss of generality, we assume trees to be binary.} using
the following operations: {\bf (1)} \emph{create\_tree}$(\gamma)$,
 which creates a tree containing only a root with label $\gamma$;
{\bf (2)} \emph{expand}$(u,\gamma_1,\gamma_2)$, which, given a leaf $u$ 
and two labels $\gamma_1$ and $\gamma_2$, creates two
new nodes (one per label) and adds them as children of $u$;
{\bf (3)} \emph{remove}$(u)$ which, given a leaf $u$ of the tree, removes it from the tree.

The environment representation problem is associated to the management of variables
and their bindings. This can be modeled as attributes of the nodes in the
tree, assuming a set of attributes $\Gamma$. At each node $u$, three operations are possible:
{\bf (1)} \emph{assign}$(\alpha,u)$ which associates the label $\alpha$ to node $u$; 
{\bf (2)} \emph{dereference}$(\alpha,u)$ which identifies the nearest ancestor $v$ of $u$ in which 
the operation \emph{assign}$(\alpha,v)$ has been performed; 
{\bf (3)} \emph{alias}$(\alpha,\beta,u)$ which requires that any \emph{dereference} operation
for $\alpha$ in any descendant of $u$ in the tree  produces the same result as the \emph{dereference} of $\beta$. 
This abstraction formalizes  the informal considerations presented in the
existing literature and can be used to classify the methods to address the 
environment representation problem~\cite{gupta-bj-crit}.

It has been demonstrated that it is impossible to derive a solution to the
environment representation problem which achieves a constant-time solution to all the operations described above.
In particular, \citeN{op-ngc} demonstrated that the overall problem has a lower-bound complexity of
$\Omega(\log n)$ on pure pointer machines, where $n$ is the maximum number of nodes appearing in the tree. The 
investigation by \citeN{op-complang} provides also an optimal theoretical solution, achieving
the  complexity of $O(\log n)$; the solution makes use of a combination of generalized linked lists, which allow
us to  insert, delete, and compare positions of nodes in the list, and AVL trees. Even though
the solution is theoretical, it suggests the possibility  of improvement over existing 
methodologies, which  have a complexity of $O(n)$.

\subsubsection{Recent Solutions to the Environment Representation Problem}

As introduced in Section~\ref{sec:review},
before 2000, the state of the art in Or-parallel Prolog systems was achieved using one of
two approaches: 
 \emph{stack copying} (as used in the Muse system and in ACE \cite{muse,annals}), and
 \emph{binding arrays} (as used in the Aurora system \cite{aurora}).  
These techniques provided comparable benefits and fostered a wealth of additional research (e.g., in the area of scheduling \cite{dutra-ilps94,warren-93,kn:Ali93}).
The advent of distributed computing
architectures, especially Beowulf clusters, led researchers to investigate the development of 
Or-parallel Prolog systems in absence of shared memory. 
The two approaches above 
are not immediately suitable for distributed memory architectures, as they require some level of 
data sharing between workers\footnote{We use the generic term \emph{worker} to denote a computing entity in
	a parallel Prolog system.} during the execution. 
For example, in the case of stack copying, 
\emph{shared frames} need to be created to coordinate workers' access to unexplored choice point alternatives. 

The PALS system \cite{iclp01,pals1,stack-split} introduced the concept of \emph{stack splitting} as a new environment
representation methodology suitable for distributed memory systems. Stack splitting replaces the
dynamic synchronization among workers, e.g., as in the access to shared choice points during backtracking
in the binding-arrays method, with a \emph{``static''} partitioning of the unexplored alternatives between
workers. Such partitioning is performed each time two workers collaborate to share unexplored branches
of the resolution tree. Thus, stack splitting allows us to remove synchronization requirements by 
preemptively partitioning   the remaining unexplored alternatives at the moment of sharing. The splitting allows both
workers to proceed, each executing its branch of the computation,
without any need for further synchronization when accessing parts of the resolution tree which are 
common among the workers.

\begin{figure}[h]
	\centerline{\fbox{\includegraphics[width=.58\textwidth]{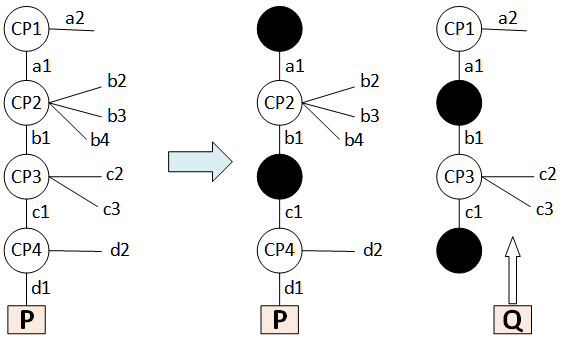}}}
	\centerline{\fbox{\includegraphics[width=.58\textwidth]{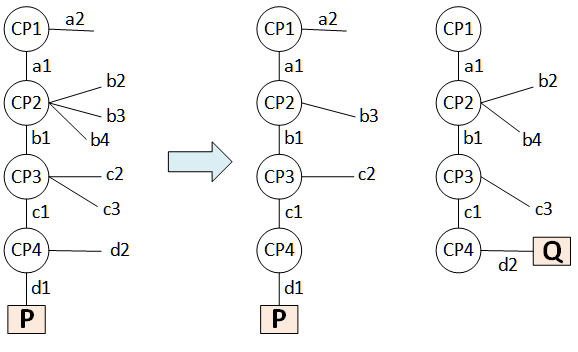}}}
	\centerline{\fbox{\includegraphics[width=.58\textwidth]{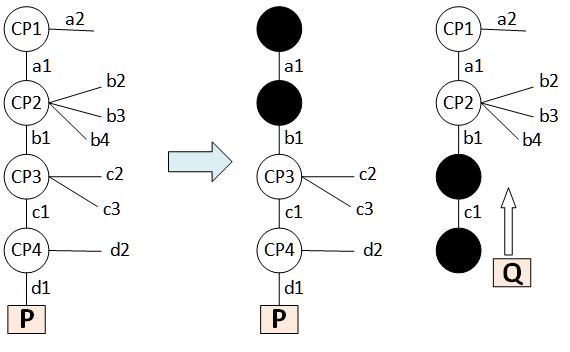}}}
	\caption{From top to bottom: Vertical Splitting,  Horizontal Splitting, Middle Splitting.
	Workers visit the resolution tree in depth first order; 
	CP stands for choice point;  CP1 is the ``root'' of the  tree. 
	P and Q denote different workers.
	A black node means that the choice point has no alternatives available. A connection between 
	P/Q and a node indicates that the worker is executing the given alternative of the choice point.
	The arrow denotes immediate
	backtracking for the worker stealing work.}
	\label{splitting_examples}
\end{figure}

The original definition of stack splitting \cite{icpp01,villa1} explored two strategies of splitting. \emph{Vertical stack splitting}
(Figure~\ref{splitting_examples}--top) is suitable for computations where branches of the resolution tree are highly unbalanced and
where choice points tend to have a small number of alternatives (e.g., binary trees). In this case,  the available choice points are 
alternated between the two sharing workers---e.g., the first worker keeps the first, third, fifth, etc. choice point with unexplored alternatives, while the second worker keeps the second, fourth, sixth, etc. In the end, each worker ends up with approximately half of the 
 choice points with unexplored alternatives. 
The alternation between choice points kept and choice points given away is meant to improve the chance of a balanced distribution of 
work between the two workers.
\emph{Horizontal stack splitting} has been, instead, designed to support sharing in the case of computations with few choice 
points, each with a large number of unexplored alternatives. In this case, the two workers  partition
 the unexplored alternatives within each available choice point (Figure~\ref{splitting_examples}--middle).

Follow-up research has explored additional variations of stack splitting. \emph{Diagonal stack splitting} \cite{diagSplit} provides a combination of horizontal and vertical stack splitting. \emph{Middle stack splitting}, also known as 
\emph{Half stack splitting} \cite{sideef}, is similar to vertical stack splitting, but with the difference that one worker keeps the top half of the choice points while the second worker keeps the bottom half. The advantage of this methodology is its efficient implementation and the ability to know the respective position of the two workers in the resolution tree immediately after the sharing operation---a useful property in order to manage side-effects and other order-sensitive predicates  (Figure \ref{splitting_examples}--bottom).

Even though stack splitting has been originally designed to sustain Or-parallelism
on distributed memory architectures, this methodology 
 provides advantages also on shared memory
platforms. 
The ACE system \cite{stack-split} demonstrated that stack-splitting
outperforms stack-copying in a variety of benchmarks. 
A microprocessor-level simulation,
using the Simics simulator, identified an improved cache behavior as the reason for the
superior performance in shared memory platforms \cite{pontellisimics}.

\subsubsection{Systems and Implementations}

The original stack splitting methodology has been developed in the context of the
PALS project---an investigation of execution of Prolog on distributed memory machines. The
PALS system introduces stack splitting by modifying the  ALS Prolog system \cite{ALS-sys}, resulting
in a highly efficient and scalable implementation. PALS explores not only novel techniques for
the environment representation problem, but also the challenges associated to management
of side-effects \cite{sideef} and different scheduling strategies \cite{villa-sched}. PALS operates on 
Beowulf clusters, mapping Prolog workers to processes on different nodes of a cluster and using 
 MPI for communication between them. PALS implements vertical, horizontal, and middle stack splitting.
 
\citeN{neworp} implemented Or-parallelism on a multi-threaded implementation of YAP Prolog
(ThOr). Their implementation relies on YapOr~\cite{Rocha-99b} 
which is based on the stack copying model.
A multi-threaded implementation can exploit low-cost parallel architectures (such as common 
 multi-core processors).

Santos and Rocha~\citeyear{SR2013,yap-or} extended this approach to
systems composed of clusters of multi-core processors---addressing the
challenge of dealing with the combination of shared and distributed
memory architectures. They propose a layered model based on \emph{two
  levels} of workers, single workers and teams of workers, and the
ability to exploit different scheduling strategies, for distributing
work among teams and among the workers inside a team. A team of
workers is formed by workers which share the same memory address space
(workers executing in different computer nodes cannot belong to the
same team). A computer node can contain more than one team.  Several
combinations of scheduling strategies are allowed. For distributed
memory clusters of multi-cores, only static stack splitting is allowed
for distributing work among teams. Inside a team, static and dynamic
scheduling can be selected. The use of static stack splitting
techniques (horizontal, vertical, diagonal, and so on) in shared
memory architectures is described by \citeN{VIEIRA2012}.  As far as
dynamic stack splitting techniques, the authors rely on the
\emph{or-frame} data structures originally introduced in the Muse
system.

\subsection{And-Parallelism}
\label{sec:andpar}


\subsubsection{Backtracking in Independent And-Parallel Prolog Systems}
\label{sec:and-backtracking}

As explained in Section~\ref{sec:review},  Independent
And-parallel (IAP) Prolog systems allow the parallel execution of subgoals
which do not interfere with each other at run time, even if they
produce multiple answers.  Section~\ref{sec:review} also pointed out
that one of the most important areas of research in the context of IAP
systems is the implementation of backtracking among parallel subgoals.
The basic issues involved in this process have been  addressed
by~\citeN{hermenegildo-phd-short} and \citeN{Backtracking}. These works identify the
main challenges of backtracking in IAP, such as trapped nondeterministic subgoals and garbage
slots. These concepts have been  further developed and improved  by
\citeN{flexmem-europar96}, 
\citeN{efficiency-complang96}, and 
\citeN{boris-phd}. 

An alternative and more efficient model for backtracking in IAP has been
proposed by~\citeN{ace-backtrack-tpds}.
The implementation model  is an extension of the original
backtracking scheme developed by Hermenegildo and
Nasr~\citeyear{Backtracking} and includes a memory organization scheme and
various optimizations to reduce communication and overhead.  It also
makes use of an adaptation of the ACE abstract interpretation-based
analyzer~\cite{dep-par-iclp}, derived from that of
the Ciao/\&-Prolog system~\cite{effofai-toplas,annotators-jlp}. In
particular, detection and special treatment of \emph{external
  variables} (i.e., variables appearing in the parallel call but created
before the call itself) enable increased independence during backtracking. 
The results obtained
show that speedups achieved during forward execution are not lost in
heavy backtracking, and the ``super-linear'' speedups that can be obtained thanks
to the semi-intelligent nature of backtracking in independent
And-parallelism are also preserved.
\footnote{This is really due to the change in the
  backtracking algorithm that independent And-parallel systems
  implement, which is a simple form of intelligent backtracking
  that takes advantage of the independence information.}


A common aspect of most  traditional IAP implementations
that support backtracking over parallel subgoals is the use of
recomputation of answers and sequential ordering of subgoals during  backtracking. While
this can, in principle, simplify the implementation, recomputation can be
inefficient if the granularity of the parallel subgoals is large 
and they produce several answers, while sequentially ordered
backtracking limits parallelism. An alternative parallel backtracking 
model has been proposed by Chico de Guzm\'an et al.~\citeyear{parback-chico-ICLP11}
which features parallel out-of-order backtracking and
relies on answer memoization to reuse and combine answers.
Whenever a parallel subgoal backtracks, its siblings also perform
backtracking after storing the bindings generated by previous
answers, which
are reinstalled when combining answers. In order  to not
penalize forward execution, non-speculative And-parallel subgoals which
have not been executed yet take precedence over sibling subgoals which
could be backtracked over.
This approach brings performance advantages and simplifies
the implementation of parallel backtracking.



The remaining challenges associated to trapped non-deterministic subgoals and garbage
slots have been addressed by~\citeANP{chico-casas-padl12}\citeyear{chico-casas-padl12}; 
the proposed
approach builds on the idea of  a
single stack reordering operation, that offers several advantages and
tradeoffs over previous proposals. While the implementation of the
stack reordering operation itself is not simple, in return it frees
the scheduler from the constraints imposed by other schemes. As a
result, the scheduler and the rest of the run-time machinery can
safely ignore the trapped subgoal and garbage slot problems and their
implementation is greatly simplified. Also, standard sequential
execution remains unaffected.

\subsubsection{High-level implementation of Unrestricted And-Parallelism} 
\label{sec:and-easy}

Original implementations of independent And-parallelism, as
demonstrated in the \&-Prolog and the \&ACE systems, relied
on low-level manipulations of the underlying abstract machine---typically
some variant of the Warren Abstract Machine. This low-level approach
is necessary in order to reduce execution overheads related to the management
of parallelism. While this approach
has been extensively used and produced highly efficient implementations, it
leads to highly complex implementations, often hard to maintain and not
able to keep up with improvements in sequential implementation technology.

\paragraph{High-level implementation.} 
\label{sec:highlevel}

The improved performance of processors and the introduction of
advanced compilation techniques have opened the doors to an
alternative approach to the implementation of parallelism---by
describing parallel execution models at a high level,  in terms
of Prolog predicates. The concept builds on the identification of a 
 small collection of built-in
predicates and their use to encode, as meta-programs, high-level 
models of parallelism
\cite{amadeo-phd,hlfullandpar-iclp2008,andprolog-new-padl2008}.
The concept of implementing a form of parallelism via meta-programming
encodings is not new (see for example the papers by
\citeN{Codish86},
\citeN{duality}, and
\citeN{ciao-ilps95}), but
it has been brought to full fruition only recently,
 in terms of a full
implementation using a minimal set of core predicates.
This is done by implementing in the engine a comparatively
small number of concurrency-related primitives which take care of
lower-level tasks, such as locking, stack set management, thread
creation and management, etc. The implementation of parallel models (e.g., 
independent And-parallelism) is  then realized through meta-programs that 
make use of such primitives.
  The approach does not eliminate altogether modifications to
the abstract machine, but it does greatly simplify and encapsulate
them, and it also facilitates experimenting with different
alternative models of parallelism.

This approach also supports the implementation of flexible solutions for some of the main
problems found in And-parallel implementations.  In fact, the
solutions presented by~\citeANP{parback-chico-ICLP11} \citeyear{parback-chico-ICLP11,chico-casas-padl12} have all been 
implemented taking advantage of the flexibility
afforded by this new approach. The experiments also show that,
although such source-level implementation of parallelism introduces 
overheads, the
performance penalties are reasonable, especially if paired with some form of
granularity control.
This is the (unrestricted) IAP implementation supported currently by
the Ciao Prolog system.

\paragraph{Unrestricted And-Parallelism.} 
\label{sec:unrestricted}

Another interesting aspect of the approach
by~\citeANP{andprolog-new-padl2008} 
is that it facilitates the implementation of alternative models of And-parallelism. In 
particular, this approach has facilitated the exploration of \emph{unrestricted And-Parallelism}, i.e.,
a form of IAP which does not follow the traditional fork-join structure used in the
previous IAP systems. This flexibility has served 
 as target for new parallelizers for unrestricted IAP
(see Section~\ref{sec:staticanal}). This work also shows that
the availability of unrestricted parallelism contributes to improved parallel
performance.

The high-level level primitives used in this system to express
unrestricted And-parallelism are those described 
in the papers by
\citeN{att-var-iclp-short},
\citeN{ciao-dis-impl-prode-short}, and
\citeN{daniel-phd}:
\begin{itemize}
\item \texttt{G \&> H} schedules subgoal \texttt{G} for parallel execution
  and continues with the code after \texttt{G}. \texttt{H} is a
  \emph{handle} that provides access to the state of subgoal \texttt{G}. 

\item \texttt{H <\&} waits for the subgoal associated with \texttt{H}
  (\texttt{G}, in the previous item) to finish. At that point, all
  bindings \texttt{G} could possibly generate are ready, since
  \texttt{G} has reached a solution. Assuming subgoal independence
  between \texttt{G} and the calls performed while \texttt{G} was
  being executed, no binding conflicts will arise. This point is also
  the ``anchor'' for any backtracking performed by \texttt{G}.
\end{itemize}
Note that, with the previous definitions, the \texttt{\&/2} operator
can be expressed as:

\centerline{\texttt{A \& B :- A \&> H, call(B), H <\&}.}

The approach shares some similarities with the 
concept of futures in parallel functional
languages~\cite{flanagan95:futures,DBLP:journals/toplas/Halstead85}. A future is meant to hold the
return value of a function so that a consumer can wait for its
complete evaluation. However, the notions of ``return value'' and
``complete evaluation'' do not make sense when logic variables are
present. Instead, \texttt{H <\&} waits for the moment when the
producer subgoal has completed execution, and the ``received values'' (typically a
tuple) will be whatever (possibly partial) instantiations have
been produced by such subgoal.

The new operators
 are very flexible, allowing the encoding of And-parallel executions which are not tied to the
 fork-join model implicit in the \& operator. In particular, one can interleave execution of 
 subgoals to ensure that variables are bound and thus enable an increasing level of parallelism. For
 example, in a computation where it has been determined (through,
 e.g., global analysis) that {\tt p(A,B)} produces the values for {\tt A,B}, {\tt q(C)} produces the
 value for {\tt C}, while {\tt r(A)} and {\tt s(C,B)} consume such values, the following encoding provides
 a higher degree of parallelism than a traditional fork-join structure:
 \begin{verbatim}
    q(C), p(A,B) &> H1, 
            r(A) &> H2,
            H1 <&,
           s(C,B), H2 <&. 
 \end{verbatim}
 The \&-Prolog engine implemented these constructs natively offering high parallel performance.
 The Ciao Prolog system includes the higher-level (Prolog) management of threads 
 \cite{hlfullandpar-iclp2008,andprolog-new-padl2008}; for example,
 the {\tt \&>} operator can be encoded as:
\begin{verbatim}
  Goal &> Handle :- add_goal(Goal, nondet, Handle),
                    undo(cancellation(Handle)),
                    release_suspended_thread.
\end{verbatim}
which adds the {\tt Goal} to a goal queue and enables one thread if available.

In closing, let us observe that these constructs are very general and
can be used to explore
a wide variety of parallelization schemes, including schemes that do
not respect the original observable semantics of Prolog.

\subsubsection{Dependent And-Parallelism}
\label{sec:dep-and-prolog}

The concept of dependent And-parallelism has not been extensively investigated
beyond the ideas already presented in the previous survey by
\citeN{parallellp-survey} and briefly reviewed in
Section~\ref{sec:review}. The
state-of-the-art at the beginning of 2000 is represented by systems like ACE with the
introduction of the filtered binding model \cite{pasco}, which supports  the dynamic
management of producer and consumer subgoals for variables shared among And-parallel subgoals. Models
like this and previously proposed ones (e.g., the DDAS model \cite{kish-jlp}) suffer from
high complexity of implementation, which complicates their maintenance and
evolution. While not further explored, we envision the extension of models like 
those proposed in Section~\ref{sec:unrestricted} to  represent a more viable approach for implementation 
of dependent And-parallelism, as hinted at already, e.g., by~\citeANP{ciao-ilps95}
\citeyear{ciao-ilps95}. 

On the other hand, the techniques for dependent And-parallelism developed for Prolog have
found use and extension in logic programming systems which have evolved from Prolog, such as 
Mercury and the EAM, as described in the following sections.

\paragraph{Mercury.}
A simplified architecture for dependent And-parallelism can be found in the parallel
implementation of Mercury \cite{conway-phd}. The proposed architecture uses a multi-threaded
implementation to create a  number of workers, which concurrently execute different 
subgoals. The implementation takes advantage of the fact that the Mercury
language requires each variable to have a single producer designated at compile time. Thus, in a
legal Mercury program, it is guaranteed that only a single point in the code will try to bind a
given free variable. As a result, producers and consumers are known at time of execution and do not 
require a complex dynamic management for shared variables. This architecture
introduces sophisticated mechanisms to support context switch of suspended
subgoals. At the language level, Mercury is extended with an \&
operator analogous to the one used in 
And-parallel Prolog systems, to identify subgoals meant for parallel execution. 

The initial parallel
implementation of Mercury restricted the parallel execution to non-communicating subgoals \cite{conway-phd}. This was later
extended to an implementation that supports communication between subgoals \cite{bone-phd}. The work by 
Bone provides a number of architectural extensions and optimizations. The system proposed by
\citeANP{bone2} \citeyear{bone1,bone2} replaces the use of a single
 centralized task queue with a collection of distributed local work queues and work-stealing
 models, similarly to other And-parallel implementations of Prolog.
This was dictated
by the need to reduce the bottleneck caused by a single-access
centralized task queue. Another
component of this system is the use of a Mercury-specific cost
analysis to detect, at compile-time, promising subgoals for parallel
execution.

\paragraph{Extended Andorra Model.}
Basic and Extended Andorra Model (EAM) are presented at the end of Section \ref{sec:review}.
In simple terms, the EAM allows subgoals to proceed concurrently as long as they are deterministic
or as long as they do not request to bind non-deterministically an
external variable (this has can also be seen as a fine-grain form of
independence). 
When non-determinism in the latter case is present, the computation can split (in a form of Or-parallelism). 

The computation in the
EAM can be described as a series of rewriting operations applied to an and-or tree. The tree includes
two types of boxes: \emph{and-boxes}, representing conjunctions of subgoals, and \emph{or-boxes}, 
representing alternative clauses for a selected literal.
The BEAM \cite{beam1,beam2} is a parallel
implementation of such a rewriting process, enhanced with a collection of control rules to enhance
efficiency (e.g., by delaying splitting, the operation which realizes Or-parallelism, until no
deterministic steps are possible).
The BEAM provides a shared-memory realization of the EAM, using
an approach called RAINBOW (Rapid And-parallelism with INdependence Built upon Or-parallel Work). The BEAM
supports independent And-parallelism enhanced with determinacy, along with different implementation models
of Or-parallelism (e.g., version vectors and binding arrays).


\subsection{Static Analysis for Parallelism} \label{sec:staticanal}


\emph{Static analysis}, generally based on
Cousots' theory of abstract interpretation~\cite{Cousot77}, is a
supporting technology in a number of areas related to parallelism in
logic programming, and especially in the process of automatic
parallelization of logic programs to exploit And-parallelism.  In
fact, logic programming pioneered the development of abstract
interpretation-based analyzers ---such as MA3 and Ms~\cite{pracabsin},
PLAI~(\citeNP{mcctr-fixpt};\citeyearNP{ai-jlp};\citeNP{anconsall-acm}), 
or  GAIA~\cite{LeCharlier94:toplas}, and the extension of this style of
analysis to
CLP/CHCs~\cite{ancons-ilps,anconsall-acm,softpe}. 
Arguably, the MA3 and PLAI parallelizers for independent
And-parallelism were the first complete, practical applications of
abstract interpretation in a working compiler (see,
e.g.,~\citeN{vanroy-survey}).

Static analysis is an area that has seen significant progress in this
period. In this section we briefly focus on the advances made in
static analysis for parallelism in logic programs since the survey
paper by~\citeN{parallellp-survey}, with a quick view on main older
results.

\paragraph{Sharing analyses.} 
One of the main
instrumental properties for automatic \mbox{(And-)}par\-al\-lel\-ization is
the safe approximation of the sharing (aliasing) patterns between
variables.  Apart from being necessary for obtaining analyses that are
at the same time precise and correct, variable sharing is a basic
component of essentially all notions of
independence~(\citeNP{sinsi-jlp,effofai-toplas,annotators-jlp,delay-popl,delay-new-ilps95};\citeyearNP{consind-toplas}).
The main abstract domains developed for detecting independence
typically seek to capture either
set-sharing~(\citeNP{abs-int-naclp89};\citeyearNP{freeness-iclp91-short};\citeyearNP{ai-jlp};\citeNP{jacobs89,shsets-jlp,HillBZ02TPLP})
or
pair-sharing~\cite{sonder86,LagoonStuckey02,setsh-flops04}. 

Results in this area 
implied the development of \emph{widenings}~\cite{Cousot77} and/or alternative
more efficient representations and abstract domain operations for
sharing domains. Widenings are used in abstract interpreta\-tion-based
analyses to support infinite abstract domains and also to reduce the
cost of complex domains, such as sharing.  In both cases the essence
of the technique is to lose precision in return for reduced analysis
time (at the limit, termination) by generalizing at some point in the
analysis to a larger abstract value. This implies arriving at fixpoints
that are less precise than the minimal fixpoint, but still safe
approximations of the concrete values.

In this line, widenings of sharing have been proposed for example
by~\citeN{Zaffanella99widening}, by performing a widening switch to a
modification of Fecht's domain~\cite{Fecht-plilp96} and incorporating
other techniques such as combining with the \texttt{Pos}
domain. \citeN{shcliques-padl06} proposed an encoding of sharing 
that represents sets of variables that have total sharing (which
gives rise to large sharing sets) in a compact way and also widens
abstract values to this representation when they are close to total
sharing. \citeN{king-lazy-set-sharing} developed a lazy technique 
that postpones computations of sharing sets until they are really
needed.  \citeN{negsharing-iclp08} proposed an encoding that uses
negative information representation techniques to switch to a dual
representation for certain variables when there is a high degree of
sharing among them.  Also,~\citeN{setsharing-with-zdd-lcpc08} proposes
and studies a representation using ZBDDs.

\paragraph{Determinism and non-failure analyses.}
Inference of determinacy~\cite{determinacy-ngc09} and
non-failure~\cite{nfplai-flops04}, including multivariant (i.e.,
context/path-sensitive) analyses, based on the PLAI framework have been investigated.
%
These analyses are instrumental in And-parallelism, since if goals can
be determined to not fail and/or be deterministic, significant
simplifications can be performed in the handling of their parallel
execution. This is due to the inherent complexity in handling
backtracking across parallel goals (Section~\ref{sec:andpar}), significant parts of which can be
avoided if such information is available.

\paragraph{Cost analysis and granularity control.}
The advances in determinism and non-failure
analyses are also useful for improving the precision of cost
analysis~(\citeNP{granularity};\citeyearNP{low-bou-sas94-short};\citeyearNP{low-bounds-ilps97}),
which is instrumental in parallelism for performing task granularity
control~\cite{granularity,granularity-jsc,pedro-phd}. Such analyses
have been extended to estimate actual execution
time~\cite{estim-exec-time-ppdp08}, rather than steps, as in previous
work, which is arguably more relevant to automatic parallelization and
granularity control.  Techniques have also been developed for fuzzy
granularity control~\cite{fuzzy-gc-icfc10}.  Another important related
line of work has been the static inference of the cost of
\emph{parallelized programs}~\cite{cost-an-par-prog-lopstr19-post}.
The base cost analyses have also been extended to be
parametric~\cite{resource-iclp07-short} (where the analysis can track
user-defined resources and compound resources) and multivariant,
formulated as an abstract domain~\cite{plai-resources-iclp14-short}.

\paragraph{Static analysis for parallelization algorithms.}
There has also been progress in the development of improved, static
analysis-based parallelization algorithms that allow automatic
parallelization for non-restricted
And-parallelism~\cite{uudg-annotators-lopstr2007,amadeo-phd} (i.e.,
not limited to fork-join structures, as supported
by Casas et al.~\citeyear{hlfullandpar-iclp2008}, see Section~\ref{sec:and-easy}).
Also, improved parallelization algorithms for non-strict
independence~\cite{sinsi-jlp} using sharing and freeness information
have been proposed~\cite{nsicond-tcs}.

In another line of work, by Vidal~\citeyear{vidal-annot-peval-tplp12-short}, a
novel technique has been presented for generating annotations for independent
And-parallelism based on partial evaluation. A partial evaluation
procedure is augmented with (run-time) groundness and variable sharing
information so that parallel conjunctions are added to the residual
clauses when the conditions for independence are met.
The results are parallel annotated programs which are shown to be
capable of  achieving interesting speedups. 

Improvements in static analysis related to scalability are directly
relevant to improving the practicality of automatic parallelization
techniques.  In this context, much work has been done in improving
these aspects for LP and CLP analyses, including combined modular and
incremental analysis, 
assertion-guided multivariant
analysis, 
and analysis of
programs with assertions and open
predicates~(\citeNP{intermod-incanal-2018-iclp-tc};\citeyearNP{guided-analysis-post-lopstr18}\citeyearNP{incanal-assrts-openpreds-lopstr19-post}).

\paragraph{Run-time checking overhead.} 
The reduction of the overhead implied by run-time tests 
(through caching
techniques~\cite{cached-rtchecks-iclp2015,nataliia-phd} and also
static analysis~\cite{optchk-journal-scp-short}, similarly
to what done in the works by \citeN{effofai-toplas} and \citeN{spec-jlp-short}),
as well as in generating
static performance guarantees for programs with run-time
checks~\cite{rtchecks-cost-2018-ppdp}, have been subject of active research.  
This work was done in the context of run-time tests for assertions but it is directly relevant
to run-time checking for conditional And-parallelism.

\paragraph{Applications to other paradigms and areas.} 
The techniques developed for LP/CLP
parallelization have been used to support the
parallelization of other paradigms~\cite{tutorial-pcjspecial}. This is
a large topic that goes beyond the scope of this survey, but examples
include the application of pair-sharing~\cite{spoto:pair_sharing} and
set-sharing~\cite{shnltau-vmcai08} to object-oriented programs,
sharing analysis of arrays, collections, and recursive
structures~\cite{Marron-sharing-paste08},
identifying logically related heap regions~\cite{marronRegions09},
identification of heap-carried
dependencies~\cite{Marron-heap-deps-lcpc08}, or context-sensitive shape
analysis~(\citeNP{Marron06};\citeyearNP{Marron08-projectextend-cc}).

As stated before, the demand for precise global analysis and
transformation stemming from automatic parallelization spurred the
development of the first abstract interpretation-based ``production''
analyzers.
Having these systems readily available led to the early realization
that analysis and transformation are very useful also 
in program verification and static debugging, as illustrated by the
pioneering Ciao Prolog
system~(\citeNP{aadebug97-informal-short};\citeNP{prog-glob-an-short}\citeyearNP{ciaopp-sas03-journal-scp-short}\citeyearNP{ciao-design-tplp-short}).
%
During the past two decades, the analysis and verification of a large
variety of other programming paradigms ---including imperative,
functional, object-oriented, and concurrent ones--- using LP/CLP-related
analysis techniques, has received significant interest, and many
verification approaches and tools have recently been implemented which
are based in one way or another in a translation into LP/CLP (referred
to in this context as Constrained Horn Clauses
--CHCs)~(\citeNP{Peralta-Gallagher-Saglam-SAS98};%
\citeNP{HGScam06-short};\citeNP{decomp-oo-prolog-lopstr07-short};%
\citeNP{NMHLFM08};\citeyearNP{resources-bytecode09};%
\citeNP{jvm-cost-esop};%
\citeNP{mod-decomp-jist09-short};\citeNP{GrebenshchikovLPR12};%
\citeNP{DBLP:conf/cav/GurfinkelKKN15-short};\citeNP{AngelisFPP15-short};\citeNP{KahsaiRSS16-short};\citeNP{resource-verification-tplp18};%
\citeNP{isa-energy-lopstr13-final};\citeyearNP{isa-vs-llvm-fopara};%
\citeNP{big-small-step-vpt2020-short}). 
%
The main reason is that LP and CLP 
are effective as languages for specifying program semantics and
program properties.
\citeNP{DeAngelisEA_TPLP21} offers an up-to-date, comprehensive survey
of this approach.
  

\subsection{Parallelism and Tabling} \label{sec:tabling}

Tabling, introduced by 
\citeNS{Chen-96},
is a powerful implementation
technique that overcomes some limitations of traditional Prolog
systems in dealing with recursion and redundant
sub-computations. Tabling has become a popular and successful
technique thanks to the ground-breaking work in the XSB Prolog system
and  in the SLG-WAM engine~\cite{Sagonas-98}. The success of SLG-WAM led to several
alternative implementations that differ in the execution rule, in the
data-structures used to implement tabling, and in the changes to the
underlying Prolog engine \cite{Swift-12,yap-journal-2012,Zhou-12,Guo-01,Somogyi-06,Chico-08,Desouter-15,Zhou-15}.

Tabling is a refinement of SLD resolution that stems from one simple
idea: programs are evaluated by saving intermediate answers for tabled
subgoals so that they can be reused when a \emph{similar call} appears
during the resolution process. First calls to tabled subgoals are
considered \emph{generators} and are evaluated as usual, using SLD
resolution, but their answers are stored in a global data space,
called the \emph{table space}. Similar calls are
called \emph{consumers} and are resolved by consuming the answers
already stored for the corresponding generator, instead of
re-evaluating them against the program clauses. During this process,
as further new answers are found, they are stored in their table
entries and later returned to all similar calls. Call similarity thus
determines if a subgoal will produce their own answers or if it will
consume answers from a generator call. 

A key procedure in tabled evaluation is the \emph{completion}
procedure, which determines whether a subgoal is \emph{completely
evaluated}. A subgoal is said to be completely evaluated when all its
possible resolutions have been performed, i.e., when no more answers
can be generated and all consumers have consumed all the available
answers. A number of subgoals may be mutually dependent,
forming a \emph{strongly connected component}
(\emph{SCC}), and therefore can only be completed
together. In this case, completion is performed by the \emph{leader}
of the SCC, which is the oldest generator subgoal in the SCC, when all
possible resolutions have been made for all subgoals in the SCC~\cite{Sagonas-98}.


\subsubsection{Parallel Tabling}


The first proposal on how to exploit implicit parallelism in tabling
is due to~\citeN{Freire-95} that has been not implemented later in
available systems.  More recent approaches are based on reordering the
execution of alternatives corresponding to multiple clauses that match
a goal in the search tree \cite{nengfa-linear-tab,Guo-01}. Guo and
Gupta's approach is a tabling scheme based on \emph{dynamic reordering
  of alternatives with variant calls}: both the answers to tabled
subgoals and the (looping) alternatives leading to variant calls are
tabled.  After exploiting all matching clauses, the subgoal enters a
looping state, where the looping alternatives start being tried
repeatedly until a fix-point is reached. The process of retrying
alternatives may cause redundant recomputations of the non-tabled
subgoals that appear in the body of a looping alternative. It may also
cause redundant consumption of answers if the body of a looping
alternative contains several variant subgoal call. Within this model,
the traditional forms of parallelism can still be exploited and the
looping alternatives can be seen as extra unexplored choice point
alternatives. However, parallelism may not come so naturally as for
SLD evaluations as, by nature, tabling implies sequentiality -- an
answer can not be consumed before being found and consuming an answer
is a way to find new ones -- which can lead to more recomputations
of the looping alternatives until reaching a fix-point.

The first system to implement support for the combination of tabling
with some form of parallelism was the YAP Prolog
system~\cite{yap-journal-2012}. A first design, named the OPTYap
design~\cite{Rocha-99a}, combines the tabling-based SLG-WAM execution
model with implicit Or-parallelism using shared memory processes. A
second design supports explicit concurrent tabled evaluation using
threads~\cite{Areias-12a}, where from the threads point of view the
tables are private, but at the engine level the tables are shared
among threads using a \emph{common table space}. To the best
of our knowledge, YAP's designs are the only effective implementations
with experimental results showing good performance on shared-memory
parallel architectures. 

The XSB system also implements some sort of explicit concurrent tabled
evaluation using threads that extends the default SLG-WAM execution
model with a \emph{shared tables design}~\cite{Marques-08}. It uses a
semi-naive approach that, when a set of subgoals computed by different
threads is mutually dependent, then a \emph{usurpation operation}
synchronizes threads and a single thread assumes the computation of
all subgoals, turning the remaining threads into consumer threads. The
design ensures the correct execution of concurrent sub-computations
but the experimental results showed some
limitations~\cite{Marques-10}.

Other proposals for concurrent tabling relies on a
distributed memory model. \citeNS{Hu-PhD} was the first to
formulate a method for distributed tabled evaluation termed
\emph{Multi-Processor SLG (SLGMP)}. As in the approach of \citeN{Freire-95},
each worker gets
a single subgoal and it is responsible for fully exploiting its search
tree and obtain the complete set of answers. One of the main
contributions of SLGMP is its controlled scheme of propagation of
subgoal dependencies in order to safely perform distributed
completion. 

A different approach for distributed tabling was proposed by~\citeNS{Damasio-00}. 
The architecture for this proposal relies
on four types of components: a \emph{goal manager} that interfaces
with the outside world; a \emph{table manager} that selects the
clients for storing tables; \emph{table storage clients} that keep the
consumers and answers of tables; and \emph{prover clients} that
perform the evaluation. An interesting aspect of this proposal is the
completion detection algorithm. It is based on a classical credit
recovery algorithm~\cite{Mattern-89} for distributed termination
detection. Dependencies among subgoals are not propagated and,
instead, a controller client, associated with each SCC, controls the
credits for its SCC and detects completion if the credits reach the
zero value. An implementation prototype has also been developed, but
further analysis is required.


\subsubsection{YAP Prolog}

We focus here on the already cited YAP Prolog system that 
provides the ground technology for
both implicit and explicit concurrent tabled evaluation, but
separately. From the user's point of view, tabling can be enabled
through the use of single directives of the form ``\emph{:-~table
p/n}'', meaning that common sub-computations for \emph{p/n} will be
synchronized and shared between workers at the engine level, i.e., at
the level of the tables where the results for such sub-computations
are stored. Implicit concurrent tabled evaluation can be triggered if
using the OPTYap design~\cite{Rocha-05a}, which exploits implicit
Or-parallelism using shared memory processes. Explicit concurrent
tabled evaluation can be triggered if using the thread-based
implementation~\cite{Areias-12a}; in this case,
the user still needs to implement the thread management and scheduler policy for
task distribution.


\paragraph{Implicit Or-Parallel Tabled Evaluation.}

The OPTYap system builds on the YapOr~\cite{Rocha-99b} and
YapTab~\cite{Rocha-00b} engines. YapOr extends YAP's sequential engine
to support implicit Or-parallelism based on the environment copying
model. YapTab extends YAP's execution model to support (sequential)
tabled evaluation. In the OPTYap design, tabling is the base component
of the system as, most of the time, each worker behaves as a full
sequential tabling engine. The Or-parallel component of the system is
triggered to allow synchronized access to the shared region of the
search space or to schedule work. Work sharing is implemented through
stack copying with \emph{incremental copying} of the
stacks~\cite{Ali-90a}.


Workers exploiting the public (shared) region of the search space must
synchronize to ensure the correctness of the tabling
operations. Synchronization is required: (i) when backtracking to
public generator or interior (non-tabled) nodes to take the next
available alternative; (ii) when backtracking to public consumer nodes
to take the next unconsumed answer; or, (iii) when inserting new
answers into the table space. Moreover, since Or-parallel systems can
execute alternatives early, the relative positions of generator and
consumer nodes are not as clear as for sequential tabling. As a
result, it is possible that generators will execute earlier, and in a
different branch than in sequential execution. Or that different
workers may execute the generator and the consumer calls. Or, in the
worst case, workers may have consumer nodes while not having the
corresponding generators in their branches. This induces complex
dependencies between workers, hence requiring a more elaborated and
specific \emph{public completion} operation~\cite{Rocha-01}.

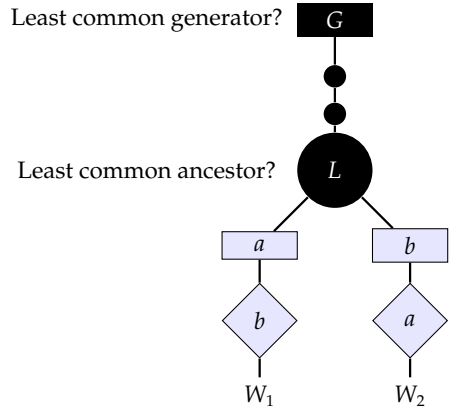
\begin{figure}

\begin{tikzpicture}[xscale=0.5,yscale=0.5]
\draw  (5,15)  node [shape=rectangle,fill=black,text=white,minimum width=1cm] (1) {$G$};
\draw  (5,13.5) node [shape=circle,fill=black] (2) {};
\draw  (5,12.5) node [shape=circle,fill=black] (3) {};
\draw  (5,11)  node [shape=circle,fill=black,text=white,minimum width=1cm] (4) {$L$};
\draw  (3,9) node [shape=rectangle,fill=blue!10,draw,minimum width=1cm] (5) {$a$};
\draw  (7,9) node [shape=rectangle,fill=blue!10,draw,minimum width=1cm] (6) {$b$};
\draw  (3,7) node [shape=rectangle,rotate=45,fill=blue!10,draw,minimum height=7mm, minimum width=7mm] (7) {};
\draw  (3,7) node  {$b$};
\draw  (7,7) node [shape=rectangle,rotate=45,fill=blue!10,draw,minimum height=7mm, minimum width=7mm] (8) {};
\draw  (7,7) node {$a$};
\draw  (3,5) node [shape=rectangle,fill=white] (9) {$W_1$};
\draw  (7,5) node [shape=rectangle,fill=white] (10) {$W_2$};

\draw  [line width=0.3mm ] (1) -- (2);
\draw  [line width=0.3mm ] (2) -- (3);
\draw  [line width=0.3mm ] (3) -- (4);
\draw  [line width=0.3mm ] (4) -- (5);
\draw  [line width=0.3mm ] (4) -- (6);
\draw  [line width=0.3mm ] (5) -- (7);
\draw  [line width=0.3mm ] (6) -- (8);
\draw  [line width=0.3mm ] (7) -- (9);
\draw  [line width=0.3mm ] (8) -- (10);


\draw  (0,15) node [shape=rectangle,fill=white] (11) {Least common generator?};
\draw  (0,11) node [shape=rectangle,fill=white] (12) {Least common ancestor?};

\end{tikzpicture}
\caption{Public completion scheme.
Black nodes are public nodes, rectangle nodes are generators, rhombus nodes are consumers. 
$W_1$ and $W_2$ are the workers.}
\label{fig_public_completion}
\end{figure}

Figure~\ref{fig_public_completion} illustrates this kind of
dependencies. Starting from a common public node, worker \texttt{W1}
takes the leftmost alternative while worker \texttt{W2} takes the
rightmost. While exploiting their alternatives, \texttt{W1} calls a
tabled subgoal \texttt{a} and \texttt{W2} calls a tabled
subgoal \texttt{b}. As this is the first call to both subgoals, a
generator node is stored for each one. Next, each worker calls the
tabled subgoal firstly called by the other, and consumer nodes are
therefore allocated. At that point, we may question at which (leader)
node should we check for completion? In OPTYap, public completion is
performed at the
least common ancestor node (node \texttt{L} in
Figure~\ref{fig_public_completion}). But that leader node can be any
type of node and not necessarily a generator node as in the case of
sequential tabling.

Consider now the case where \texttt{W1} has explored all its private
work and backtracks to the public leader node \texttt{L} common
to \texttt{W2}. Since work is going on below \texttt{L}, \texttt{W1}
cannot complete its current SCC (which includes \texttt{L}, the
generator node for \texttt{a} and the consumer node
for~\texttt{b}). The reason for this is that \texttt{W2} can still
influence \texttt{W1}'s branch, for instance, by finding new answers
for subgoal \texttt{b}. On the other hand, we would like to
move \texttt{W1} in the tree, say to node \texttt{N}, where there is
available work and, for that, we may need to reset the stacks to the
values in \texttt{N}. As a result, in order to allow \texttt{W1} to
continue execution, it becomes necessary to \emph{suspend the SCC} at
hand. This is the only case where Or-parallelism and tabling conflict
in OPTYap (a worker needs to move in the tree above an uncompleted
leader node). OPTYap's solution is to save the SCC's stacks to a
proper space, leaving in \texttt{L} a reference to where the stacks
were saved. These suspended computations are considered again when the
remaining workers check for completion at \texttt{L}. To resume a
suspended SCC a worker needs to copy the saved stacks to the correct
position in its own stacks, and thus, it has to suspend its current
SCC first. To minimize that, OPTYap adopts the strategy of resuming
suspended SCCs \emph{only when the worker finds itself at a leader
node}, since this is a decision point where the worker either
completes or suspends its current SCC. OPTYap's public completion
algorithm and associated data structures is one of the major
contributions of OPTYap's design.


\paragraph{Explicit Concurrent Tabled Evaluation.}

In YAP, threads run independently within their own execution
stacks. For tabling, this means that each thread evaluation depends
only on the computations being performed by itself, i.e., from the
thread point of view, each thread has its own private tables but, at
the engine level, YAP uses a \emph{common table space} shared among
all threads.

\begin{figure}
\begin{tikzpicture}[xscale=0.5,yscale=0.5]
\draw  (5,17.5)  node [shape=rectangle, minimum height=1cm,minimum width=3cm,draw, rounded corners=5pt] 
     (1) {$\begin{array}{c}\mbox{Tabled Predicate}\\ \mbox{compiled code} \end{array}$};
\draw  (5,15)  node [shape=rectangle,minimum width=3cm,minimum height=1cm,draw,fill=gray,text=white] 
     (2) {\bf Table Entry};
\draw  (5,12.5) node [shape=rectangle,	minimum width=6cm,minimum height=1cm,draw, rounded corners=5pt] 
    (3) {Subgoal Trie Structure};
\draw  (-2,9.5)  node [shape=rectangle,  draw, rounded corners=5pt] 
    (4) {$\begin{array}{c}\rm Subgoal \\ \rm Frame \\ call_1\end{array}$};
\draw  (3,9.5)  node [shape=rectangle,  draw, rounded corners=5pt] 
     (5) {$\begin{array}{c}\rm Subgoal \\ \rm Frame \\ call_2\end{array}$};
\draw  (7,9.5) node {$\dots$};
\draw  (11,9.5)  node [shape=rectangle,  draw, rounded corners=5pt] 
     (6) {$\begin{array}{c}\rm Subgoal \\ \rm Frame \\ call_n\end{array}$};
\draw  (-2,6.5)  node [shape=rectangle,  draw, rounded corners=5pt] 
     (7) {$\begin{array}{c}\rm Answer \\ \rm Trie \\ \rm Structure\end{array}$};
\draw  (3,6.5)  node [shape=rectangle,  draw, rounded corners=5pt] 
     (8) {$\begin{array}{c} \rm Answer \\ \rm Trie \\  \rm Structure\end{array}$};
\draw  (7,6.5)  node {$\dots$};
\draw  (11,6.5)  node [shape=rectangle, draw, rounded corners=5pt] 
     (9) {$\begin{array}{c}\rm Answer \\ \rm Trie \\ \rm Structure\end{array}$};

\draw  [line width=0.3mm ] (1) -- (2);
\draw  [line width=0.3mm ] (2) -- (3);
\draw  [line width=0.3mm ] (3) -- (4);
\draw  [line width=0.3mm ] (3) -- (5);
\draw  [line width=0.3mm ] (3) -- (6);
\draw  [line width=0.3mm ] (4) -- (7);
\draw  [line width=0.3mm ] (5) -- (8);
\draw  [line width=0.3mm ] (6) -- (9);
\end{tikzpicture}

\caption{YAP's table space organization}
\label{fig_table_space}
\end{figure}
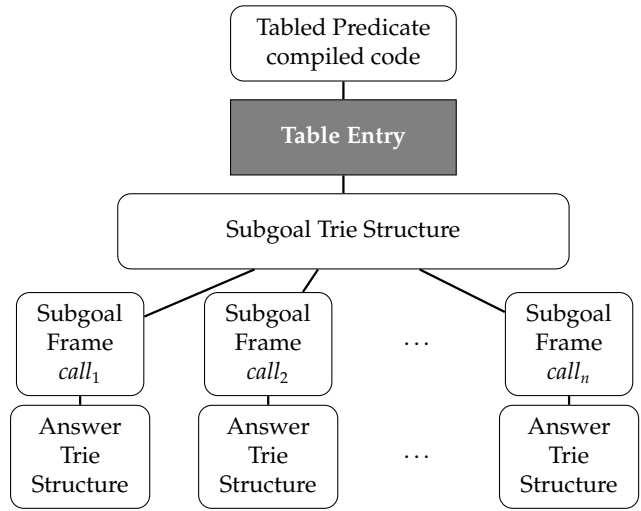

Figure~\ref{fig_table_space} shows the general table space
organization for a tabled predicate in YAP. At the entry level is
the \emph{table entry} data structure where the common information for
the predicate is stored. This structure is allocated when a tabled
predicate is being compiled, so that a pointer to the table entry can
be included in the compiled code. This guarantees that further calls
to the predicate will access the table space starting from the same
point. Below the table entry, is the \emph{subgoal trie
structure}. Each different tabled subgoal call to the predicate at
hand corresponds to a unique path through the subgoal trie structure,
always starting from the table entry, passing by several subgoal trie
data units, the \emph{subgoal trie nodes}, and reaching a leaf data
structure, the \emph{subgoal frame}. The subgoal frame stores
additional information about the subgoal and acts like an entry point
to the \emph{answer trie structure}. Each unique path through the
answer trie data units, the \emph{answer trie nodes}, corresponds to a
different answer to the entry subgoal. To deal with multithreading
tabling, YAP implements several designs with different degrees of
sharing of the table space data structures.

We report here the main features of three sharing designs proposed in~\cite{Areias-12a}:
\emph{No-Sharing (NS)}, \emph{Subgoal-Sharing (SS)}, and \emph{Full-Sharing (FS)}.

NS was
the starting design for multithreading tabling support in YAP:
each thread allocates fully private tables and the table entry
is extended with a \emph{thread array}, where each thread has its own
entry, which then points to the private subgoal tries, subgoal frames
and answer tries for the thread.

In the SS design, the
threads share part of the table space, namely, the subgoal trie
structures are now shared among the threads and the leaf data
structure in each subgoal trie path, instead of referring a subgoal
frame, it now points to a thread array. Each entry in this array then
points to private subgoal frames and answer trie structures. In this
design, concurrency among threads is restricted to the allocation of
trie nodes on the subgoal trie structures. Tabled answers are still
stored in the private answer trie structures of each
thread. The \emph{Partial Answer Sharing (PAS)}
design~\cite{areias-jss16} extends the SS design to allow threads to
share answers. The idea is as follows: whenever a thread calls a new
tabled subgoal, first it searches the table space to lookup if any
other thread has already computed the full set of answers for that
subgoal. If so, then the thread reuses the available answers, thus
avoiding recomputing the subgoal call from scratch. Otherwise, it
computes the subgoal itself. The first thread completing a subgoal
call, shares the results by making them available (public) to the
other threads. The PAS design avoids the usage of the thread array and
instead it uses a list of private subgoal frames corresponding to the
threads evaluating the subgoal call. In order to find the subgoal
frame corresponding to a thread, we may have to pay an extra cost for
navigating in the list but, once a subgoal frame is completed and made
public, its access is immediate since it is moved to the beginning of
the list.

Finally, the FS design 
tries to maximize the amount of data
structures being shared. In this design, the answer tries and part of
the subgoal frame information are also shared among threads. A
new \emph{subgoal entry} data structure stores the shared information
for the subgoal, which includes access to the shared answer trie
and to the thread array that keeps pointing to the subgoal frames,
where the remaining private information is kept. In this design,
concurrency among threads includes the access to the new subgoal
entries and the allocation of tries nodes on the answer trie
structures. Memory usage is reduced to a minimum and, since threads
share the answer tries, answers inserted by a thread for a particular
subgoal call are automatically made available to all other threads
when they call the same subgoal. However, since different threads can
be inserting answers in the same answer trie, when an answer already
exists, it is not possible to determine if the answer is new or
repeated for a particular thread without further support. This can be
a problem if the tabling engine implements a \emph{batched scheduling}
strategy~\cite{Freire-96}. To mitigate this problem, the \emph{Private
Answer Chaining (PAC)} design~\cite{areias-slate15-post} extends the
FS design to keep track of the answers that were already found and
propagated per thread and subgoal call.


\subsubsection{Perspective on the Future}

We believe that a challenging goal for the combination of tabling with
parallelism is the design of a framework that integrates both implicit
and explicit concurrent tabled evaluation, as described earlier,
 in a single tabling
engine. This is a very complex task since we need to combine the
explicit control required to launch, assign and schedule tasks to
workers, with the built-in mechanisms for handling tabling and
implicit concurrency, which cannot be controlled by the user.

In such a framework, a program begins as a single worker that executes
sequentially until reaching a \emph{parallel construct}. A parallel
construct can then be used to trigger implicit or explicit concurrent
tabled evaluation. If the parallel construct identifies
a request for  \emph{explicit concurrent
evaluation}, the execution model launches a set of
additional workers to exploit concurrently a set of independent
sub-computations (which may include tabled and non-tabled
predicates). From the workers' point of view, each concurrent
sub-computation computes its tables but, at the implementation level,
the tables can be shared following YAP's design presented
above. Otherwise, if the construct requires  \emph{implicit concurrent evaluation}, 
the execution model launches a set of additional workers
to exploit in parallel a common sub-computation. Parallel execution is
then handled implicitly by the execution model taking into account
possible directive restrictions. For example, we may have directives
to define the number of workers, the scheduling strategy to be used,
load balancing policies, etc. By taking advantage of these parallel
constructs, a user can write parallel logic programs from scratch or
parallelize existing sequential programs, by incrementally pinpointing
the sub-computations that can benefit from parallelism, using the
available directives to test and fine tune the program in order to
achieve the best performance.
Combining the inherent implicit parallelism of Prolog with
high-level parallel constructs will clearly enhance the expressiveness
and declarative style of tabling and simplify concurrent programming.



\section{Parallelism and Answer Set Programming}\label{sec:asp}

Answer Set Programming is a programming paradigm for knowledge representation and reasoning based on some
key points: the use of negation as failure, the semantics based on stable models (also known as answer sets), and a bottom-up model 
computation on  a ground version of the program. 
Many solvers for Answer Set Programming became available in the last decades. 
The work by Gebser et~al.~\citeyear{DBLP:conf/ijcai/GebserLMPRS18} is a recent survey on ASP systems.

The presentation in this section starts with a quick review of parallelism in Datalog, the  language for deductive Databases. Even though pure Datalog lacks negation (which is a crucial starting point for ASP) and uses implementation techniques which are different from ASP, we opt to start our conversation from Datalog as the original logic programming paradigm based on a bottom-up model computation---i.e., a paradigm where the result of the computation is the desired (i.e., minimal) model of a logic program. 
We then  focus on extraction of parallelism from the ASP computation; in particular, we consider parallelization of the search process used by ASP, parallelization of the grounding phase, the exploitation of portfolio parallelism, and conclude with some final considerations on other opportunities for parallelism in ASP. 
For other details on this topic the reader is
referred to the  survey by~\citeN{DBLP:books/sp/18/DovierFP18}.
Various forms of parallelism have been implemented in modern ASP solvers and experimented with
in ASP Competitions \cite{DBLP:journals/tplp/GebserMR20}.

\subsection{Parallelism and Datalog}
\label{sec:datalog}


Research on  parallelization of Datalog inherited results and techniques developed
in the field of relational DBMS, such as  parallelization of relational operations and 
of SQL query evaluation.
Particularly relevant are the approaches to parallelization of natural join 
operations, as they are at  the core of the naive and semi-naive bottom-up computation in 
Datalog,
and query optimization; the literature has explored such issues in the context of
a variety of computing  architectures~\cite{033_WangXLSLG18,044_DiamosWWLY13,070_Zeuch18,103_ZinnWWAY16,030_ShehabAS17}.

Initial attempts towards the parallelization of Datalog  appeared early in literature.
Among many, we mention the works by Wolfson and Silberschatz~\citeyear{dbase},
Ganguly et al.~\citeyear{ganguly2}, and  Zhang et al.~\citeyear{ZhangWhangChau95}
which explore the execution of Datalog on 
distributed memory  architectures.
These approaches are mainly restricted to definite programs
or, in the case of programs with negation, to stratified programs.
The core idea consists in parallelizing the computation of the minimum model of a
Datalog program, computed using
 the semi-naive bottom-up technique.
Program rules are partitioned and assigned to the distributed workers.
Communication between workers is implemented through explicit message passing. The different
early proposals differ on the techniques used to distribute sets of rules different workers
and management of the communication, used to exchange components of the minimal model
as it is computed.

Similar approaches have been developed to 
operate on multi-core shared-memory machines---by exploring hash
functions to partition computation of relations that guarantee the avoidance of locks~\cite{zaniolo15}.
Significant speedups can be obtained by using as little synchronization as possible during the program evaluation. 
An example is the work by \citeN{DUTRA13}, where the 
load of computing the model is distributed among the various GPU threads that can access and modify the 
data in the GPU shared memory.

In more recent years,
various tools and systems have been developed to evaluate Datalog programs in parallel or distributed settings.
Some of such parallel/distributed engines have been mainly designed 
to support declarative reasoning in application domains like declarative networking,
program analysis, distributed social networking, security, and graph/data analytic.
These approaches often extend plain Datalog with some form of aggregation and negation to meet 
the needs of the specific application domain.

\citeN{009_MoustafaPYD16} propose \emph{Datalography,} a bottom-up evaluation engine for Datalog tailored to graph analytics.
The system enables the distributed evaluation of Datalog
queries, also involving aggregates; the target architecture for this
implementation is  a BSP-style graph processing engine.
Queries are compiled into sets of rules that can be executed locally by the distributed workers.
Optimizations are applied to partition the work and to minimize
communication between workers. 

A different approach to large-scale Datalog applications is adopted in the \emph{Souffl\'e} system \cite{1.012_JordanSS16,012_NappaZSS19,025b_ZhaoSS20}. Efficient Datalog evaluation is obtained by
compiling declarative specifications into C++ programs involving openMP directives that enable parallel execution
on shared-memory multi-core architectures.

The \emph{BigDatalog} system \cite{023_CondieDISYZ18,026_DasZ19,085_ShkapskyYICCZ16}
supports scalable analytics and can be executed on distributed and multi-core architectures.
The notion of \emph{``pre-mappability''}~\cite{022_ZanioloYDSCI17} 
is exploited to 
extend Datalog's fixpoint semantics in order to enable aggregates in recursion.  

\citeN{R000_SeoPSL13} describe \emph{Distributed SociaLite,} an extension of the SociaLite system 
for parallel and distributed large-scale graph analysis.
In this case, users can specify how data should be partitioned and shared across workers 
in a cluster of multi-cores. The system optimizes the communication needed between workers.

In the \emph{Myria} system \cite{017_WangBH15} for large-scale data analytics,
Datalog queries, possibly involving aggregates and recursion, are translated into 
parallel query plans to be executed in a shared-nothing multi-core cluster.
Both synchronous and asynchronous evaluation schemes are possible.

Concerning the exploitation of parallelism in Datalog-based approaches to data analysis, we also mention
\emph{RecStep}~\cite{024_FanZZAKP19}, an implementation of a general-purpose Datalog engine built on top of a  
parallel RDBMS.
The target architecture is a single-node multi-core system and the language extension of plain Datalog
offers aggregates and stratified negation.
\emph{Yedalog}~\cite{YedalogChin15} extends Datalog to enable computations and analysis on  large collections 
of semi-structured data. The  \emph{DeALS} system (see the paper by \citeN{Shkapsky16thesis} and the references therein) which
supports standard SQL-like aggregates as well as user-defined aggregates,
combined with (stratified) negation, has been ported to both multicore platforms and distributed
memory systems (using the Spark library). 

In Sections \ref{sec:bigdata} and \ref{sec:gpus} we will also show other approaches to
Datalog parallelism.


\subsection{Search (Or-) Parallelism in ASP}
\label{sec:asppar}

The most popular ASP solvers proposed in the literature implement
search processes that explore the search space of possible
truth value assignments to the atoms of the ground program---directly or indirectly
(e.g., through activation of program rules). This has prompted the study
of parallelization of the search process, by supporting the concurrent exploration
of different parts of the search space. This form of parallelization resembles
the exploitation of Or-Parallelism in Prolog, as discussed earlier. 

\subsubsection{General Design}
The concept of Or-parallelism (or Search parallelism) in ASP  emerged early on in the history
of the paradigm---the first popular ASP solvers appeared around 1997
(e.g., Smodels \cite{smodelsLPNMR}, DLV \cite{dlvsystem}) and the first reports of parallel
ASP solvers were presented in 2001
\cite{workshop-stable,stab-kentucky,ipps-stable}.
These first Or-parallel ASP  implementations  originated from 
modifications of the Smodels inference engine. The intuitive structure of the algorithm underlying 
Smodels is illustrated in Algorithm \ref{smo}. 
The procedure incrementally constructs a partial interpretation
 by identifying atoms as being true (added to $S^+$) or false (added to $S^-$). 
If all atoms of the Herbrand base of the program $P$ ($B_P$) are assigned,
the solution is returned.
The \textit{Expand}
procedure is used to \emph{deterministically} expand the partial interpretation constructed so far, using the clauses of the program to infer the truth value of other atoms. Intuitively, if $\langle S^+,S^-\rangle$ is the current partial interpretation, the \textit{Expand} procedure determines a 
subset  of
$$\langle \{A\:|\: P \cup S^+ \cup \{\neg B\:|\: B\in S^-\} \models A\},
              \{A\:|\: P \cup S^+ \cup \{\neg B\:|\: B\in S^-\} \models \neg A\}\rangle$$
The implementation of \textit{Expand} in Smodels uses inference rules which are equivalent to the
computation of the  well-founded model of the program $P\cup S^+\cup \{\leftarrow p\:|\:p\in S^-\}\}$
(similar to the program transformations proposed by
Brass et al.~\citeyear{DBLP:journals/tplp/BrassDFZ01}). The \textit{Select\_Atom} function
 heuristically selects an unassigned atom to assign next, while the \textit{Choose} function creates a 
 non-deterministic choice between the two following alternatives. Or-parallelism arises from the concurrent exploration
 of the alternatives generated by \textit{Choose}---see also Figure~\ref{or-asp}.

\SetKwFor{Loop}{loop forever}{}{end}
\begin{center}
\begin{minipage}[c]{.5\textwidth}
\begin{algorithm}[H]
	\DontPrintSemicolon
	\KwInput{A ground program $P$}
	\KwOutput{Answer Set}
	
	$\langle S^+,S^-\rangle = \langle \emptyset, \emptyset\rangle$\;
	\Loop{}{
		$\langle S^+,S^-\rangle = \textit{Expand}(P,\langle S^+,S^-\rangle)$\;
		\If{$(S^+ \cap S^- \neq \emptyset)$}
		{
			\Return{{\bf Fail}}\;
		}
		\If{$(S^+\cup S^- = B_P)$}
		{
			\Return{$\langle S^+,S^-\rangle$}\;
		}
		$p = \textit{Select\_Atom}(P,\langle S^+,S^-\rangle)$\;
		\textit{Choose:}\;
		\hspace{1cm} 1: $S^+ = S^+ \cup \{p\}$\;
		\hspace{1cm} 2: $S^- = S^- \cup \{p\}$\;
	}

\caption{Intuition of Smodels.}\label{smo}
\end{algorithm}
\end{minipage}\hspace{.2cm}\begin{minipage}[c]{.45\textwidth}

\begin{tikzpicture}[xscale=0.5,yscale=0.5]
\draw  (5,18)  node []  (1) {};
\draw  (5,15.5)  node  [shape=rectangle,minimum width=3cm,minimum height=1cm,draw]  
     (2) {$\langle S^+, S^- \rangle$};
\draw (6.5,17) node [] {$Expand$};
\draw  (5,12.5) node [shape=rectangle,rotate=45,draw,minimum height=10mm, minimum width=10mm] (3) {};
\draw (7,14) node [] {$Select\_Atom$};
\draw (5,12.5) node []  {$p$};
\draw  (1,10)  node [] (4) {};
\fill [fill=gray!10, draw] (1,10) -- (-1,6) -- (3,6) -- (1,10) ;
\draw  (1,12)  node  []  {$\langle S^+ \cup \{p\}, S^- \rangle$};

\draw  (9,10)  node [] (5) {};
\fill [fill=gray!30, draw] (9,10) -- (7,6) -- (11,6) -- (9,10);
\draw  (9,12)  node  []  {$\langle S^+, S^-  \cup \{p\} \rangle$};

\draw (1,11) node [] {$Expand$};
\draw (9,11) node [] {$Expand$};

\draw  [line width=0.3mm ] (1) -- (2);
 
\draw  [line width=0.3mm ] (2) -- (3);
\draw  [line width=0.3mm, -> ] (3) -- (4);
\draw  [line width=0.3mm, ->  ] (3) -- (5);
\draw (5,10) -- (5,4);

\draw  (1,4.5)  node [] {Worker A};
\draw  (9,4.5)  node [] {Worker B};

\end{tikzpicture}

		
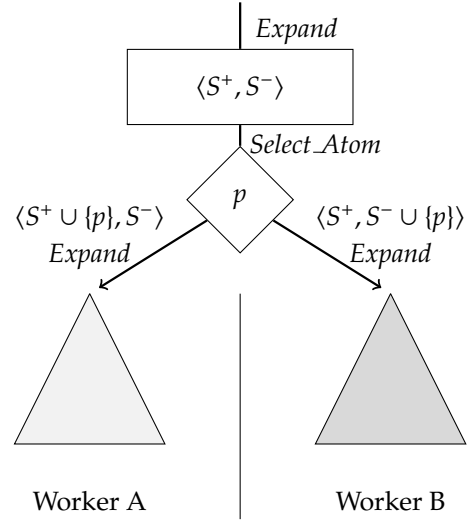
\captionof{figure}{Intuition of Or-Parallelism in ASP}
		\label{or-asp}
\end{minipage}
\end{center}

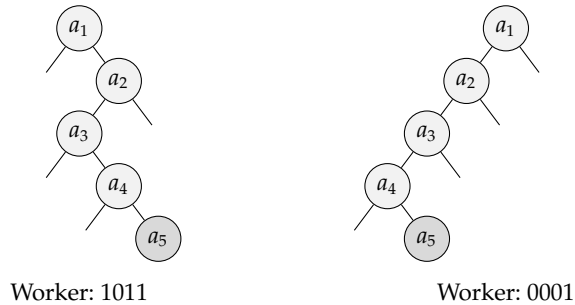
\begin{figure}[htbp]
	\begin{tabular}{ccc}
	
	\begin{tikzpicture}[xscale=0.7,yscale=0.7,level distance=10mm]
\draw (1,6) node [circle,minimum size=6mm,inner sep=1pt, draw,fill = gray!10] {$a_1$}
child {node {}}
child {node [circle,minimum size=6mm,inner sep=1pt, draw,fill = gray!10] {$a_2$}
        child { node [circle,minimum size=6mm,inner sep=1pt, draw,fill = gray!10] {$a_3$}
                      child {node{}}
                      child {node [circle,minimum size=6mm,inner sep=1pt, draw,fill = gray!10] {$a_4$}
                               child {node{}}
                               child {node [circle,minimum size=6mm,inner sep=1pt, draw,fill = gray!30] {$a_5$}}  
                              }
                 }
        child {node {}}
       }
;

\draw (1,1) node [] {Worker: 1011};
\end{tikzpicture}
& \phantom{aaaaaaaa} &
\begin{tikzpicture}[xscale=0.7,yscale=0.7,level distance=10mm]
\draw (1,6) 
node [circle,minimum size=6mm,inner sep=1pt, draw,fill = gray!10] {$a_1$}
child {   node [circle,minimum size=6mm,inner sep=1pt, draw,fill = gray!10] {$a_2$}
            child { 
                  node [circle,minimum size=6mm,inner sep=1pt, draw,fill = gray!10] {$a_3$}                     
                  child { node [circle,minimum size=6mm,inner sep=1pt, draw,fill = gray!10] {$a_4$}
                            child {node{}}
                            child {node [circle,minimum size=6mm,inner sep=1pt, draw,fill = gray!30] {$a_5$}} 
                           }
                  child {node {}}         
                   }
          child {node {}}
          }
child {node{}}                 
;

\draw (1,1) node [] {Worker: 0001};
\end{tikzpicture}
	\end{tabular}
	\caption{Examples of initial distribution in ParStab: 0 left child, 1 right child}
	\label{stab}
\end{figure}

The model proposed by \citeN{stab-kentucky} relies on a centralized scheduling structure,
with a single central master and a collection of slave workers. The master is responsible
for coordinating the distribution of unexplored parts of the search tree among workers. Paths
in the search tree are described as binary strings (i.e., a $0$ represent a left child while a
$1$ represents a right child). Initially, each worker receives the complete ASP program and a 
binary string which is used to deterministically choose its position in the search tree (Figure
\ref{stab}). A similar model has been adapted for execution on distributed memory architectures in 
the claspar system \cite{claspar0,claspar2}. The claspar system adds the ability of organizing 
workers in a deeper hierarchical structure and the ability to exchange learned nogoods across 
parallel computations. 

The binary search tree created by the execution of Smodels is irregular---thus leading to
an unbalanced distribution of work among workers. When a 
worker has exhausted its assigned search tree, it requests a new branch to the master; in turn,
the master requests work from a randomly chosen worker. The selected worker will 
transfer the highest (i.e., closest to the root) choice point with open alternatives
to the master for redistribution, marking it as fully explored. This approach avoids the risk
of different workers exploring the same branch of the search tree. Selecting unexplored
choices closer to the root is a known heuristic aimed at increasing the chance of 
assigning to a worker a potentially large task.

The system initially proposed by \citeN{workshop-stable} and \citeN{ipps-stable}, and later fully
developed by  \citeN{parasp} and \citeN{parasp1}, represents a fully Or-parallel implementation of Smodels
with symmetric workers---without the presence of a master serving as broker for distribution of 
unexplored tasks. Each worker explores parts of the search tree as well as participates in the 
distribution of unexplored parts of the search tree to other, idle, workers. Thus, each worker
alternates {\bf (1)}~\emph{computation} steps (corresponding to the execution of Smodels), and 
{\bf (2)}~\emph{load balancing} steps to relocate workers to branches of the search tree with
unexplored alternatives. The design has been developed on both shared memory systems
\cite{ipps-stable,parasp} as well as on Beowulf clusters \cite{parasp1}. The two implementations
have a similar design, where branches of the search tree are represented locally
within the data structures of  each worker (i.e., a complete Smodels solver).

\subsubsection{Scheduling and Heuristics}
The process of load-balancing is essential to allow workers to remain busy and increase the
degree of parallelism exploited. Load balancing is composed of two activities:
{\bf (1)}~\emph{scheduling,} which is used to identify the location in the search tree where an idle
worker should be moved, and {\bf (2)}~\emph{task sharing,} which is the actual process of 
moving the worker to the new location.

\begin{figure}[htbp]
	\includegraphics[width=.55\textwidth]{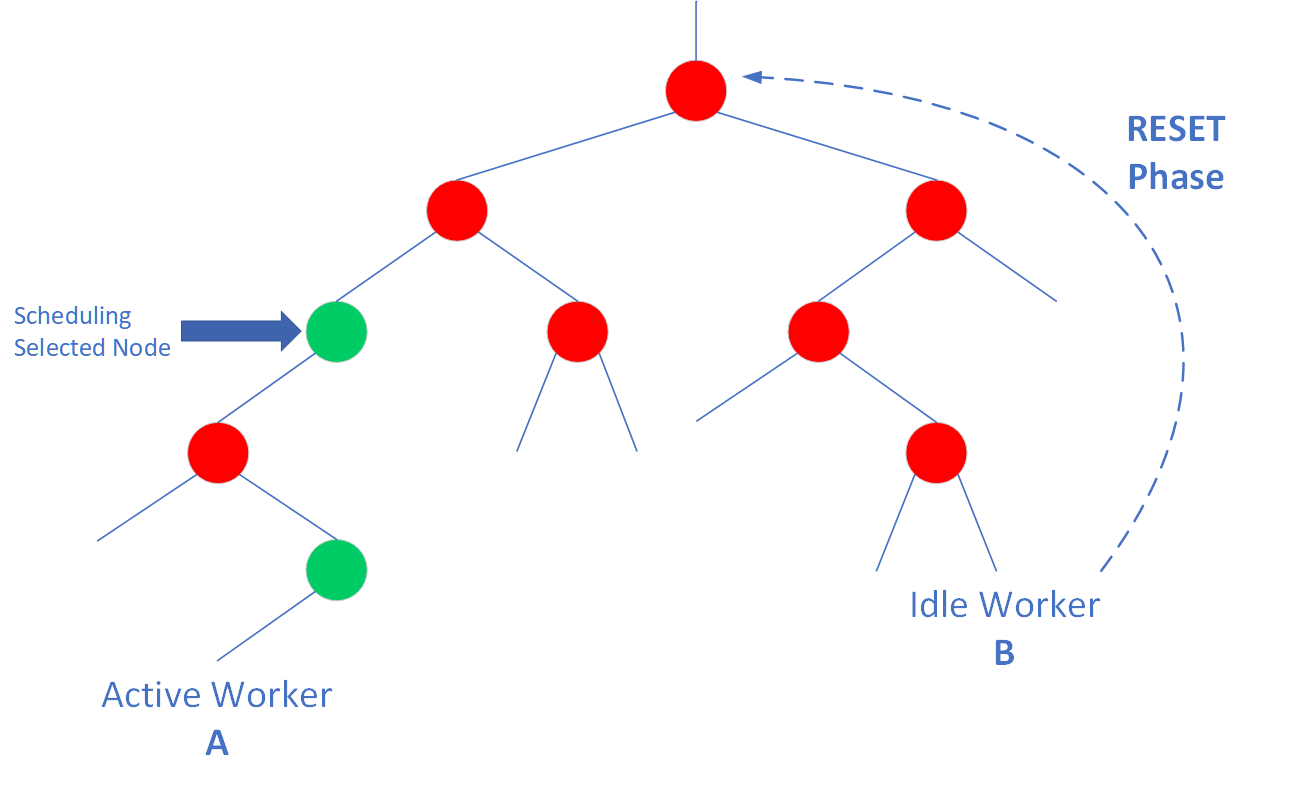}
	\includegraphics[width=.44\textwidth]{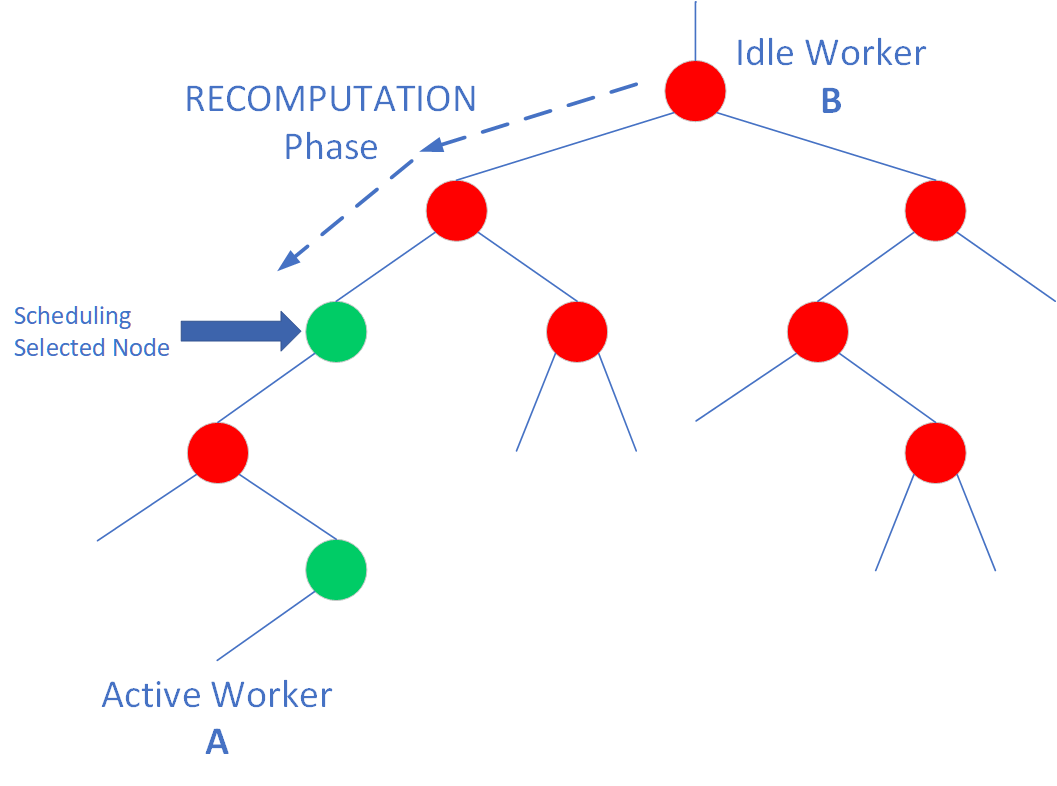}
	\includegraphics[width=.43\textwidth]{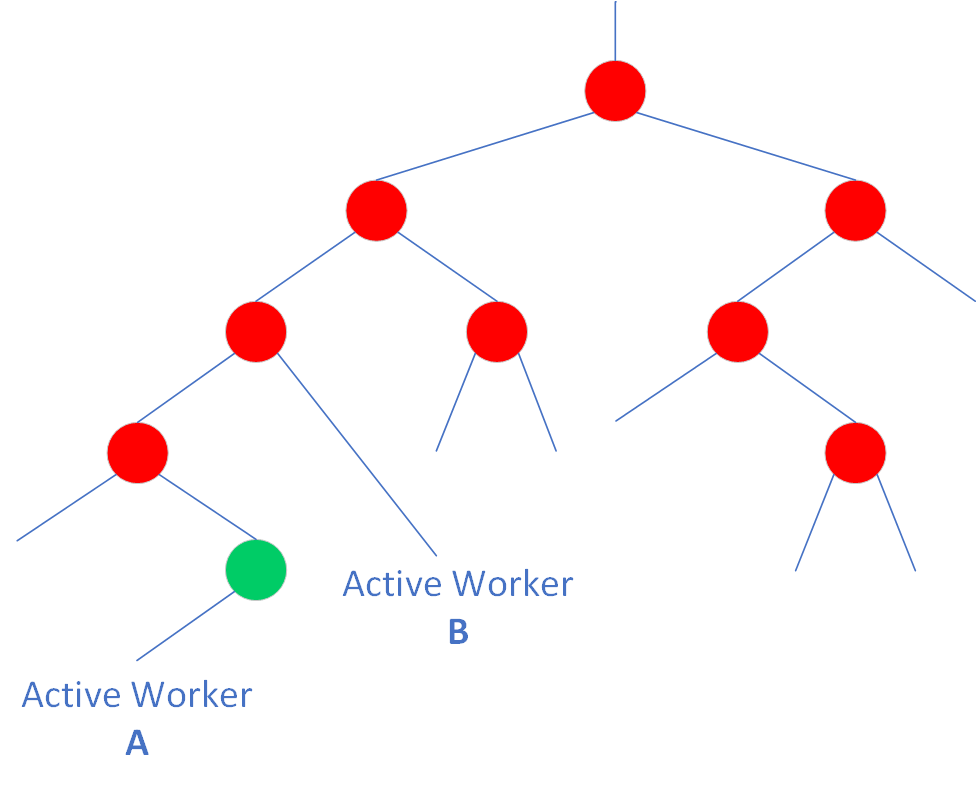}
	\caption{Recomputation with Reset}
	\label{recomp}
\end{figure}

Several dimensions have been explored for both scheduling and task sharing, and experimental
comparisons have been presented by Le et al.~\citeyear{le1,le07,parasp1}. 
While significant performance differences 
can be observed, thus making it hard to identify a clear winning strategy, on average the most 
effective methodology for scheduling and task sharing is the \emph{Recomputation with Reset} 
strategy (Figure \ref{recomp}). Intuitively, scheduling is based on selecting the highest node
in the tree with unexplored alternatives. The process of task sharing is realized by restoring 
the state of the worker to the root of the tree (Figure \ref{recomp} top left) and repeating the computation from the
root to the selected node of the tree (Figure \ref{recomp} top right). The latter operation can be performed efficiently using
a single \emph{Expand} operation. The idle worker is then able to restart the computation with an unexplored alternative from the selected node (Figure \ref{recomp} bottom). The idea of recomputation in Or-parallelism is not novel---it has been explored in the context of Or-parallelism in Prolog by several systems, such as the Delphi model \cite{delphimodel} and the Randomized Parallel Backtracking model 
\cite{randpar1,randpar2}.

\subsection{Parallel Grounding}
\label{sec:parground}

In this section we focus on the parallelization of the grounding stage that transforms the first order logic program $P$ into an equivalent propositional  program $ground(P)$.  $P$ uses a finite set of constant
symbols $\mathcal{C}$. A rule using $n$ different variables should, 
in principle, be replaced by $|\mathcal{C}|^n$ ground clauses where
variables are replaced by elements of  $\mathcal{C}$ in all possible ways.
This simple idea can be easily parallelized; however 
it might lead to a ground program of unacceptable size.
Since the first grounder Lparse \cite{DBLP:journals/ai/SimonsNS02}
this problem has been addressed by
splitting program-defined predicates into two classes: domain and non-domain
predicates. The precise definition of domain predicates has been changed in the
evolution of grounders, but the idea is that they are those predicates that 
can be extensionally computed deterministically using a bottom-up procedure.
In particular, all predicates extensionally  defined by ground facts
are domain predicates.
Every variable occurring in a rule should occur in the argument of 
a positive atom in the body as well to help grounding (domain/range restriction).
This introduces a partial ordering between parts of the program.
This ordering has been exploited by parallel grounders, as well.
Precisely,  parallelism for grounding can be implemented at three 
different levels:

\begin{enumerate}
	\item \emph{Components Level} parallelism is based on the analysis of the strongly connected components (SCC) of the dependency graph $\mathcal{G}(P)$.
The program is split in modules and the grounding follows the topological ordering of the SCC. Independent modules can be managed in parallel and synchronization points are used to control the overall process.

\item
\emph{Rules Level} parallelism. Each rule can be, in principle, grounded in parallel. 
Non-recursive rules are grounded first.
Grounding of rules involved in recursion are delayed. 
Their grounding follows the bottom-up fixpoint procedure (precisely, the  
{semi-naive} evaluation procedure developed for Datalog) 
that ends when no new ground clauses are generated.

\item
\emph{Single Rule Level} parallelism takes care of parallelizing into different threads the grounding of a single rule.
Assume a rule contains $n$ different variables $X_1,\dots,X_n$. 
Each variable $X_i$, $i \in \{1,\dots,n\}$, 
occurs in an atom based on a domain predicate and we know that $X_i$ ranges in 
the set of constant symbols  $\mathcal{C}_i \subseteq \mathcal{C}$.
The  grounding of the rule produces $\Pi_{i=1}^n |\mathcal{C}_i|$ ground instances.
It is expected that $|\mathcal{C}_i| \ll |\mathcal{C}|$ but, in any case, the number of instances 
grows exponentially with $n$.
Thus, implementing this form of parallelism is crucial for rules containing several variables.

\end{enumerate}

The research in parallel grounding can be traced back to the work of Balduccini et al.~\citeyear{parasp},
where authors exploited the property of range restrictedness
of  Lparse programs and implemented single rule parallelization
following a static partition of rules. Let us observe
that each parallel processor is assumed to be aware of the domain predicates used in the
rule. The group of University of Calabria has deeply investigated the 
multilevel parallelism hierarchy described above (see, for instance, the work by
\citeN{DBLP:journals/jal/CalimeriPR08} and by \citeN{DBLP:journals/tplp/PerriRS13}) with outstanding performance 
improvements. An interesting observation of these works is the evidence that, in the majority of the explored
benchmarks, single rule level parallelism represents the dominating component for performance improvement.

\subsection{Other Forms of Parallelism}
\label{otherASP}
 
Alternative forms of parallelism have also been explored in the
context of ASP. 

Lookahead parallelism has been considered as a technique to improve
performance of the deterministic steps of the ASP computation. Lookahead
is an optimization technique introduced in the Smodels system. Lookahead
is part of the \emph{Select\_Atom} operation. 
Before selecting a chosen atom, the lookahead operation performs 
a quick set of \emph{Expand} operations (which are typically very
efficient) on undefined atoms: given a partial answer set
$\langle S^+, S^-\rangle$ and a set of atoms $A$ such that
$A \cap (S^+\cup S^-)=\emptyset$,  
for each $x\in A$ the system executes both \emph{Expand}$(P,\langle S^+\cup\{x\},S^-\rangle)$
and \emph{Expand}$(P, \langle S^+, S^-\cup\{x\}\rangle)$. If both of them leads to failure, then
backtracking should be initiated; if only one of them leads to failure, then $x$ can be added
to the partial answer set and the process continue; if both operations succeed without failure, then 
the element can be considered as an option for the non-deterministic choice. The intuition of
parallel lookahead is to perform such tests concurrently to eliminate unsuitable options for 
choice. \citeN{parasp} demonstrate speedups in the range from 5 to 24 using 50 processors
on a variety of benchmarks.

\subsection{Portfolio Parallelism}
\label{sec:portfolio}


ASP solvers, constraint solvers (used in CLP), and SAT solvers (used as auxiliary solvers in logic programming languages such as Picat \cite{DBLP:conf/padl/ZhouK16}) are composed of many distinguished parts, but they share a common scheme: {\bf(1)} choice of a not yet instantiated variable/literal, {\bf(2)} choice of the value (e.g., true/false for ASP and SAT solvers, a domain element in many constraint solvers), {\bf(3)} deterministic propagation of the value chosen that allows us to reduce the remaining part of the search space. When the assignments made so far cause a \emph{conflict} (e.g., a rule/clause/constraint unsatisfied), a {\bf(4)} backtracking/backjumping activity starts, possibly after a (deterministic) analysis of the conflict that might lead to {\bf(5)} learning new clauses/constraints.
These clauses are implied by the problem and therefore from a logic point of view they are redundant, but their explicit addition can speed up the remaining part of the search, thanks to the new inference power they support. Moreover, the search can be sometimes \emph{restarted}  {\bf(6)} from the beginning but new parts of the search tree are visited thanks to the new constraints.

This general scheme supports many variants, especially for the choices {\bf(1)}, {\bf(2)}, and {\bf(6)}.
Thus, one could use different solvers (a \emph{portfolio of algorithms} \cite{DBLP:journals/ai/GomesS01}) or a solver with many parameters that can, in principle, be tuned to the particular instance of the problem to provide the best possible performance (\emph{algorithm configuration}).
A solver has a set of parameters with values in discrete or continuous domains. Even if continuous domains are discretized, the number of possible tuples of  parameter values 
would make a manual optimization  infeasible. The algorithm configuration approach applies statistical techniques to automatically find an ``optimal'' configuration for a family of problem instances that follow a certain distribution.  This is usually implemented by iterating a local search routine starting from an initial solution and verifying it in a set of training instances.

An easy way of exploiting parallelism in this context could be simply that of using a set of (fixed) algorithms and run them in parallel, and taking the first solution generated by one of the parallel threads. In the case of algorithm configuration, local search techniques can be run in parallel (with different random choices) and the best solution in a finite amount of time is retrieved.
Thus, algorithm portfolio and configuration activities will benefit from a parallel architecture.

The research in the area has been rather active in the last years showing excellent performances. The heuristics that drive to the choices {\bf(1)} and {\bf(2)} can be static (computed only at the beginning of the search) or dynamic (updated during the search) and based on the analysis of some features. 
Typical features are of the following types  (as described by Maratea et al.~\citeyear{DBLP:journals/tplp/MarateaPR14}):
\begin{description}
\item[Problem size features]
number of rules $r$, number of atoms $a$, ratios $r/a$
and ratios reciprocal $a/r$ of the ground program.
\item[Balance features]
ratio of positive and negative atoms in each rule body,
ratio of positive and negative occurrences of each variable,
fraction of unary, binary, and ternary rules, etc
\item[Proximity to Horn features] fraction of Horn rules and number of 
atoms occurrences in Horn rules. 
\item[ASP peculiar features]
number of true and disjunctive facts, fraction of normal rules and constraints, head sizes, occurrences of each atom in heads, bodies and rules, occurrences of true negated atoms in heads, bodies and rules, sizes of the SCCs of $\mathcal{G}(P)$, 
number of Head-Cycle Free (HCF) and non-HCF, etc
\end{description}
Moreover, dynamic features can consider the ratio of the variables currently assigned by propagation vs those assigned 
non-deterministically, the number of restarts, 
the number of rules learned since the last restart, and so on.
Modern solvers 
offer a number of built-in sets of parameter configurations.
For instance,  the heuristics {\tt frumpy} of the ASP solver clasp~\cite{DBLP:journals/aicom/GebserKKOSS11} sets the following parameters
\begin{verbatim}
 --eq=5 --heuristic=Berkmin --restarts=x,100,1.5 
 --deletion=basic,75 --del-init=3.0,200,40000 
 --del-max=400000 --contraction=250 
 --loops=common  --save-p=180 --del-grow=1.1 
 --strengthen=local --sign-def-disj=pos
\end{verbatim}
In particular, it uses the variable selection rule developed for the SAT solver
{\tt Berkmin} \cite{DBLP:journals/dam/GoldbergN07}.
Furthermore, clasp allows the user to choose the  {\tt auto} option that select the most promising configuration based on the features of the current instance or to input a precise configuration from a file. 

The selection of the algorithm  is implemented using supervised and unsupervised machine learning techniques. In the case of algorithm configuration, instead, parameters are often tuned using local search technique. We give a brief summary here of the approaches in the area of SAT, ASP, and Constraint Programming (a complete and up-to-date survey can be found in the paper by \citeN{DBLP:conf/cilc/Tarzariol19}).

SATzilla \cite{SATZILLA} is the first algorithm selection implementation in the area of
SAT solving. Several versions follow its first prototype adding new statistical techniques and obtaining much more performances.
CLASPfolio \cite{DBLP:conf/lpnmr/GebserKKSSZ11} executes
 algorithm selection among a set of twelve configurations of the clasp solver 
 with ``complementary strengths'' using support vector regression techniques. 
A static approach is made by ME--ASP \cite{DBLP:journals/tplp/MarateaPR14} analyzing a  set of parameters as explained above.

A common experience in problem-solving is that (often) a solver either solves a problem in few seconds or does not solve it in days. Then, the idea is to use the algorithm portfolio with a limited time and analyze the output for selecting the best algorithm and repeat this several times  during the search. This technique is called \emph{algorithm scheduling}.
In this case, it is crucial that the portfolio selection is made in parallel.
This technique has been implemented and applied with success in Constraint Programming  by
SUNNY ---\emph{SUbset of portfolio solvers by using k-Nearest Neighbor technique to define a lazY learning model}---\cite{DBLP:journals/tplp/AmadiniGM14}, in ASP by aspeed~\cite{DBLP:journals/tplp/HoosKLS15} that is based on the clasp solver, and in SAT by \citeN{DBLP:conf/cp/MalitskySSS12}. All the three proposals exploit multi-core architectures and parallelism.

Portfolio parallelism has been considered  in the second revision 
of claspar~\cite{claspar2}. Portfolio parallelism is
realized by instructing a pool of workers to attempt to solve the same problem but with different
configurations of the solver---e.g., different heuristics and different parameters to guide search. This allows
the workers to ``compete'' in the resolution of a problem by creating different organizations of the
search space and exploring them in parallel. \citeN{claspar2} provided speedups of the order of 2
using this technique.

Some systems  combine algorithm portfolio and configuration.
CLASPfolio 2 \cite{DBLP:journals/tplp/HoosLS14}
improves the selection technique of CLASPfolio by defining a static pre-solving scheduling that may intervene if the learned selection model performs poorly.
AutoFolio \cite{DBLP:journals/jair/LindauerHHS15} executes 
Algorithm Configuration over CLASPfolio 2, choosing the optimal learning techniques with its optimal parameters configuration.

\smallskip

As a final observation, the use of portfolio techniques may have an impact on the ``observable semantics'' of a system in some
situations. In the context of  solving optimization problems or in determining all models of a program, portfolio techniques will
not modify the behavior of the system; on the other hand, if the system is used to determine one answer set, then portfolio techniques may lead to a different answer set than the one found by a sequential system.


\section{Going Large: Logic Programming and Big Data Frameworks}
\label{sec:bigdata}


The scalability of logic programming technologies has been a constant
focus of attention for the research community. Scalability in terms
of speed has been well understood and materialized in a number of highly
efficient systems. Scalability in terms of memory consumption, instead, is still
an open challenge.
There are several examples offered in the literature that capture this challenge. For example,
\begin{itemize}
	\item In the domain of planning using ASP, scalability in terms of solving large
	planning problem instances is negatively impacted by the large grounding produced by the
	combination of the number of actions and plan length. For example, in the encoding of the popular
	Biochemical Pathway planning benchmarks, from the International Planning Competitions, 
	ASP can ground only instances with less than 70 actions (instances 1--4), running out of memory
	with Instance 5 (which contains 163 actions) \cite{SonP07}.
	\item The use of logic programming  techniques for processing knowledge bases (e.g., RDF stores)
	faces the need for smart preprocessing techniques or interfaces with external databases in order
	to cope with the sheer size of large repositories---e.g., as in the case of the CDAOStore, 
	an ASP-based system for phylogenetic inference over a repository with 
	5 Billion triples, stored in 957~GB \cite{cdaostore,iclp12}.
\end{itemize}
These challenges have prompted a number of research directions, ranging from
the use of interfaces between logic programming systems and external repositories (e.g., 
to avoid the need for fully in-memory reasoning), as in the DLV-HEX system \cite{dlvhex}, to 
systems working with lazy-grounding or non-ground computations~\cite{bonatti2,gasp,lazy1}.
An alternative approach relies on the use of distributed programming techniques to 
address scalability.

\subsection{Introduction to Large Scale Data Paradigms}
The literature has offered access to popular infrastructures and paradigms to facilitate
the development of applications on distributed platforms, taking advantage of both the parallelism
and the ability to distribute data over the memory hierarchies of multiple machines. In this
section, we will briefly review some of the fundamentals of such distributed programming
infrastructures (see, e.g., Hadoop and Spark documentation \cite{hadoop1,hadoop2} for further insights).

\paragraph{Distributed File Systems (DFS)}
are designed to provide a scalable and highly fault-tolerant file system, which can be deployed on a collection
of machines using possibly low-cost hardware. The \emph{Hadoop Distributed File System (HDFS)}
represents a popular implementation of a DFS \cite{hdfs}. It adopts a
master-slave architecture. \emph{NameNodes} (masters) regulate access to files, track data files,
and store  metadata. File content is partitioned  into small \emph{chunks} (blocks) of data
distributed across the network. Chunks may range in size (typically, from $16~MB$ to $64~MB$)
 and replicated to provide redundancy and seamless recovery in case of hardware
 failures. An HDFS instance may consist of thousands
 of server machines and provides hardware failure detection, a write-once-read-many access
 model, and streaming data access.

\begin{figure}[htbp]
\begin{center}
	\begin{minipage}[t]{.45\textwidth}
	\fbox{\includegraphics[width=\textwidth]{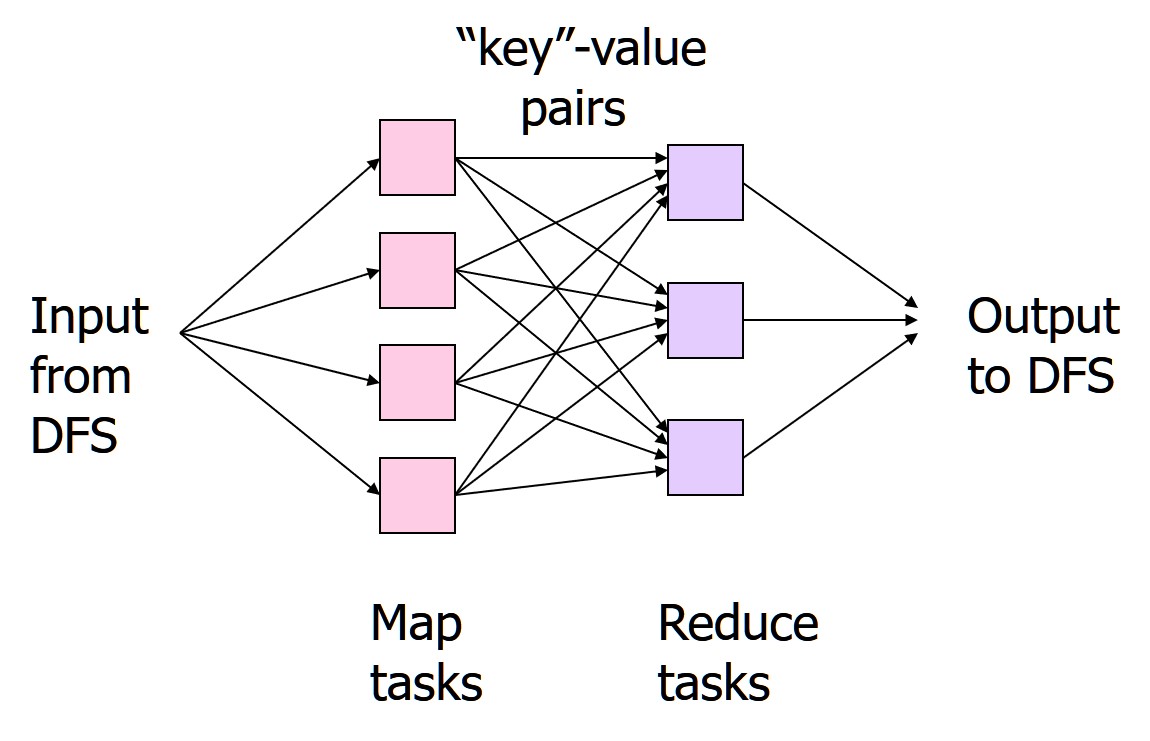}}
	\end{minipage}
	\hspace{.5cm}
	\begin{minipage}[t]{.45\textwidth}
		\fbox{\includegraphics[width=\textwidth]{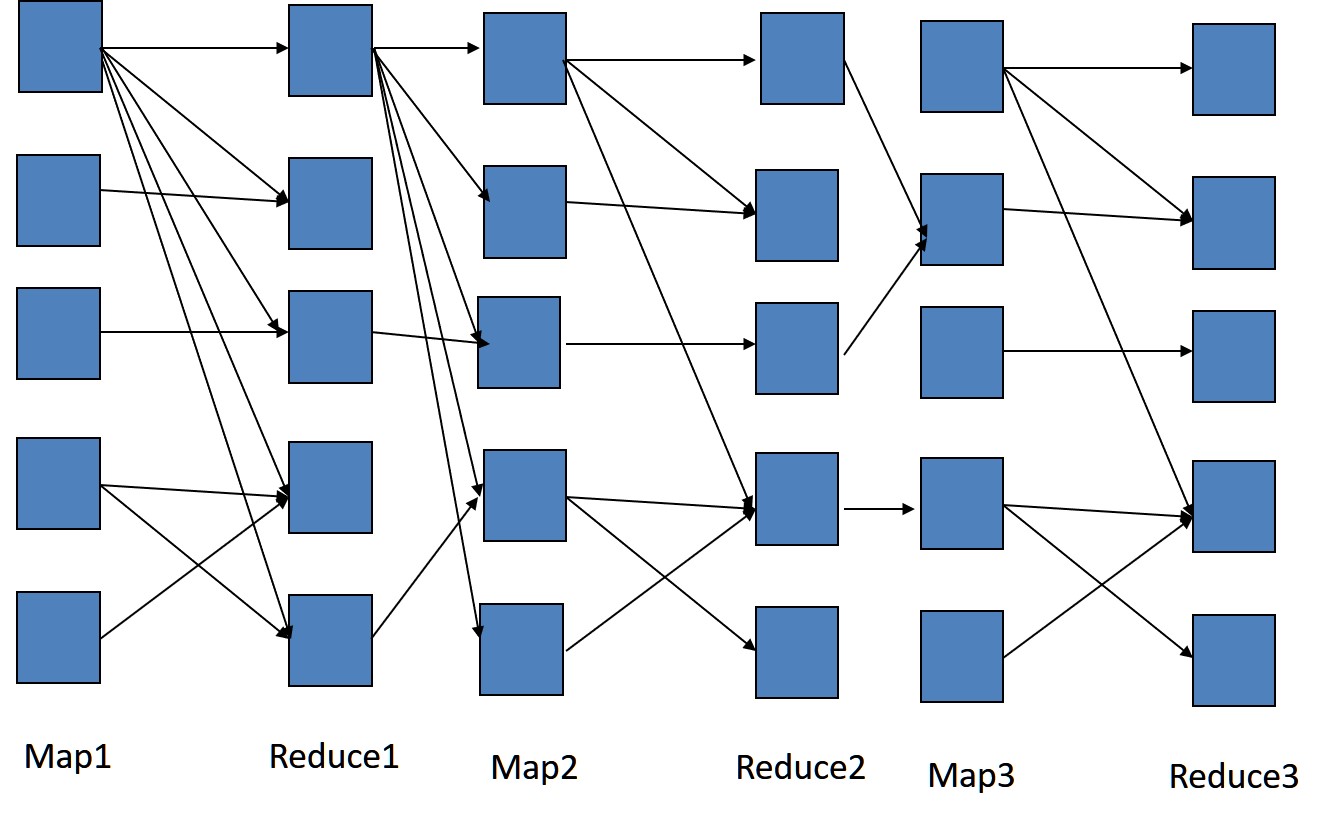}}
	\end{minipage}
\end{center}	
\caption{Map-Reduce paradigm}
\label{mr}
\end{figure}

\paragraph{Map-Reduce}
is a distributed programming paradigm designed to analyze and process large
data sets. 
The foundations of the concept of Map-Reduce can be traced back to 
fundamental list operations in functional programming. 
However, the concept gained popularity as
a rigid paradigm for  distributed programming, piloted by Google and popularized in implementations
as in the Apache Hadoop framework \cite{mapred,hdfs}. 
The Map-Reduce paradigm provides a basic
interface consisting of two methods (see Figure~\ref{mr}--left):
\begin{itemize}
	\item {\tt map( )} that maps a function over a collection of objects. It outputs a collection of ``key-value''	 	tuples;
	 \item {\tt reduce( )} that takes as input a collection of key-value pairs and merges the values of all
	  entries with the same key. 
\end{itemize}
The map method is usually performed in order to transform and filter a collection, while 
the reduce method usually performs a summary (e.g., counting the elements of a collection, or the word
 frequencies in a text).

Extended implementations of Map-Reduce have been devised to allow the iterative execution (Figure~\ref{mr}--right) of Map-Reduce cycles
optimizing communication \cite{haloop,twister}.

\paragraph{Graphs Primitives} are made available for HDFS within the framework 
Apache Spark \cite{hadoop2}.  
Spark is an in-memory data processing engine that allows for streaming,
data processing, machine learning and SQL functionalities; it relies on the concept of 
 \emph{Resilient Distributed Dataset  (RDD).} 
 A RDD is an immutable, fault-tolerant distributed collection of objects, a read-only,
partitioned collection of records, organized into logical partitions, that may be located and
 processed on different nodes of the HDFS.
 Among the libraries built on top of Spark Core, \emph{GraphX} \cite{graphx} has been developed for graphs
  modeling and graph-parallel computation. GraphX takes full advantage of the RDD data
  structure and extends it providing a distributed property multigraph abstraction. Property
  graphs are  directed multigraphs with user-defined objects associated to vertices and
  edges. They are encoded by pairs of RDDs containing the properties for vertices and edges,
  and, therefore, inherit RDDs features, such as map, filter, and reduce.

 GraphX also gives access to a complete interface for dealing with graphs, as well to
  an implementation of the Pregel API \cite{MalewiczABDHLC10}.
  Pregel is a programming model for large-scale graph
  problems and for fix-point computations over graphs. A typical Pregel computation
  consists of a sequence of \emph{super-steps.} Within each super-step, vertices may interact with their
  neighbors by sending messages. Each vertex analyzes the set of messages received in the
  previous super-step (if any) and alters its content according to a user-defined function. In turn,
  each node can generate new messages to the neighboring nodes. A Pregel computation stops when
  a super-step does not generate any messages or when other halting conditions are
  encountered (e.g., a maximum number of iterations).

\subsection{Large Scale Computing in Datalog}

\paragraph{Computing Natural Joins Using Map-Reduce.}

The underlying core component of most implementations of logic programming
using Map-Reduce is the ability to scale the computation of natural join operations
over large datasets, as originally studied by Afrati et al.~\citeyear{mapreduce2,mapreduce1}.

Let us start considering the basic case of a 2-way join between two relations $R$ and $S$,
denoted by $R \lrtimes S$.
Let us assume, for the sake of simplicity, that $R$ and $S$ are binary and that the second attribute name of the former is also the 
first attribute name of the latter (briefly, it can be denoted as $R(A,B) \lrtimes S(B,C)$).
A Map-Reduce computation can be achieved
as follows:
\begin{itemize}
	\item Each Map task receives tuples drawn from the two relations $R$ and
		$S$ and produces key-values pairs where the $key$ is the value of the
		common argument $B$ and the values are the remaining component of the tuple.
	      Namely, $(a,b) \in R$  is mapped to the pair $\langle b, (a,R)\rangle$, and
	      $(b,c) \in S$ is mapped to $\langle b, (c,S)\rangle$
	      ($R$ and $S$ are here the names of the two relations).

	\item Each Reduce task receives a pair with \emph{one value} of the common argument and 
		a list of all tuples containing such value, e.g., 
		$\langle b,L\rangle$, with $L= [(a,R),(a',R),(c,S),(c',S),\dots]$ and outputs a
		resulting (ternary) relation $RS$ with tuples $(x,b,y) \in RS$ for each
		$(x,R),(y,S)$ in the list $L$.
\end{itemize}
Thus, the Reduce task exploits data parallelism, using one worker for every singe element of the common domain.

This can be extended to the case of multi-way joins, e.g., $R(A,B)\lrtimes S(B,C) \lrtimes T(C,D)$, without
the need of cascading two-way joins (which could lead to very large intermediate relations). The model
explored relies on the use of hashing. Let us adopt a hash function~$h$ for the attribute $B$, partitioned
in $\beta$ buckets, and a hash function~$g$ for the common attribute $C$, partitioned in $\gamma$ buckets.
\begin{itemize}
	\item Each Map task performs the following hash operations:
		{\bf (1)} each tuple $(x,y) \in R$  is mapped to
			$\langle (h(y),c), (R,x,y)\rangle$ for each $1 \leq c \leq \gamma$;
		{\bf (2)} each tuple $(y,z) \in S$ is mapped to
			$\langle (h(y),g(z)), (S,y,z) \rangle$; 
		{\bf (3)} each tuple $(z,w) \in T$ is mapped to 
			$\langle (b,g(z)), (T,z,w)\rangle$ for each $1\leq b \leq \beta$.
	\item For each pair $(b,c) \in \{1,\dots,\beta\} \times \{1,\dots,\gamma\}$ there is a reduce task taking care 
	       of pairs of the form $\langle (b,c), L\rangle$. 
	       It outputs the relation $RST$ with tuples $(x,y,z,w)$ for every triple of triples 
	       $(R,x,y), (S,y,z), (T,z,w)$ in $L$.
\end{itemize}
In this case the distribution of the work is split in $\beta \gamma$ workers, and $\beta$ and $\gamma$ can be defined arbitrarily, thus determining the amount of parallelism we would like to exploit.


\paragraph{Distributed Computation in Datalog.}

Distributing the computing of the join is the basis for the general computation of bottom-up semantics.
In particular, the WebPIE system \cite{webpie} has been designed to support reasoning over RDF stores. The proposed system is capable of determining inferences according to RDFS semantics and the OWL \emph{ter Horst} fragment \cite{Horst05}. The approach adopted consists of encoding the inference rules capturing these semantics as Datalog rules. 
The computation of the least fixpoint of the set of rules given a collection of RDF triples is achieved
as an iterated Map-Reduce computation, which captures the bottom-up application of the inference rules.
The WebPIE system takes advantage of the specific format of the resulting Datalog rules. In the case of RDFS,
the system takes advantage of the following factors:
\begin{itemize}
	\item Each rule has at most two subgoals in the body---thus allowing the use of the Map-Reduce method
	to compute 2-way natural joins to capture the bottom-up application of each rule;
	\item In most rules, one of the two subgoals is one of the triples from the RDF store---allowing to optimize
	the computation by keeping the original triples in memory and matching them with generated ones;
	\item In most cases, it is possible to order the application of the rules to perform the computation using
	only three phases of Map-Reduce.
\end{itemize}

The generalization to arbitrary Datalog programs has been investigated in a variety
of works (e.g., those by \citeN{mapreduce1} and by \citeN{ullman12}). While different approaches have provided a variety of 
optimizations, based on special types of rules (e.g., transitive closures) and size of the definitions of
different predicates in the extensional database, they all build on the principles of iterated execution
of Map-Reduce workflows, using approaches like HaLoop \cite{haloop}.

This work direction has been expanded to consider Datalog with a stratified use of
negation, 
as a natural iteration of the approaches described earlier \cite{mapreduce3}.
Following the lexicographical sort of the $\mathcal{G}(P)$,
the computation can proceed by iterating the standard Datalog computation (implementing the $T_P$ using 
joins and exploiting Map-Reduce).
The extra ingredient here is handling of negated literals where \emph{anti-join} needs to be implemented.

For example, given a rule of the type
$$h(A,B) \texttt{:-} p(A,C), q(C,B), \texttt{not } r(B).$$
the inference is realized using two Map-Reduce processes:
\begin{enumerate}
	\item The first implements a natural join to derive the consequences of $p(A,C),q(C,B)$, as discussed earlier (e.g., producing a temporary relation $pq(A,C,B)$);
	\item The second Map-Reduce phase performs an \emph{anti-join}---with analogous structure as the natural join, with the exception that the reduce step is used to filter out those tuples with matching values in the $pq$ relation and the $r$ relation.
\end{enumerate}
Strong performance
results have been presented, thanks also to a broad range of optimizations based on special
cases present in the program clauses, such as lack of common arguments in the body of a rule 
\cite{mapreduce3}.

The approach can be  extended to support the computation of the well-founded semantics 
of logic programs with negation (c.f. Section \ref{sub:prel:lp}), as demonstrated by~\citeN{revisedwfs}.

Recent approaches for increasing the speedup of Datalog distributed
computation introduce new data structures
\cite{DBLP:conf/ppopp/JordanSZS19} and exploit
network topology 
\cite{DBLP:conf/cidr/BlanasKS20}
for increasing the speedup of Datalog distributed computation.

\medskip

A similar methodology was also used to support inferences in stratified Datalog with defeasible reasoning
\cite{distrdef}. In simplified terms, a theory is composed of a Datalog program with two types of rules:
$$\begin{array}{lcr}
\underset{\textit{non-defeasible}}{\textit{head} \leftarrow \textit{body}} &\phantom{aaaaaaa}&
	\underset{\textit{defeasible}}{\textit{head} \Leftarrow \textit{body}}
		\end{array}$$
along with an ordering among defeasible rules. The work by Tachmazidis et al.~\citeyear{distrdef} explores the use
of Map-Reduce in the case of stratified defeasible theories---i.e., theories where the dependency graph is 
acyclic---and thus organized in strata (e.g., the top stratum includes predicates with no outgoing edges, the
preceding one contains the predicates with links to the top stratum, etc.).
The computation requires two Map-Reduce phases for each stratum, starting with the lowest one. The first
Map-Reduce is used to identify applicable rules, using multi-way joins as discussed earlier, recording for each
derivable element the producing rule and the defeasible nature of the derivation. The second phase, which is 
primarily composed of a Reduce step, compares derived rules based on the ordering of defeasible rules to 
identify undefeated conclusions. 

\subsection{Large Scale Computing and ASP}

The use of Map-Reduce and related paradigms to support the execution of 
ASP has been only recently considered and with only very preliminary results. The 
extension of the previously mentioned approaches to the case of ASP is not trivial,
due to the switch from a bottom-up computation leading to a single minimal model, as in 
Datalog, to the case of a non-deterministic computation leading to possibly multiple
models.

The foundation of the methodology considered for the distributed computation of ASP lies in the
ability of modeling the computation of the semantics of logic programming in terms of operations of
graphs \cite{mr_asp}, and taking advantage of existing models for large scale
distributed computations over graphs. This idea was piloted by \citeN{maiterth}. 

Basically, from a ground logic program $P$,  a graph akin to the 
rule dependency graph $\mathcal{G}(P)$ is computed. 
The idea is of looking for a coloring of the nodes with two colors; colors encode the fact that
the body of a rule is \emph{activated} (by the tentative model) and therefore its head should be in the model, as well,   or not. 
A correspondence between these colorings and answer sets has been established and therefore the computation is delegated to graph operations that can be parallelized exploiting the library GraphX of Apache.

These concepts have been used to develop a preliminary ASP solver, using the Pregel support of Apache Spark \cite{igne,DBLP:conf/cilc/BortoliITDP19}. The approach allows dealing with programs with huge grounding that could not be stored in a single memory. However, answer set computation based on coloring can not exploit the many heuristics and the conflict driven clause learning techniques commonly implemented in ASP solvers and the running time is not comparable.
Map-Reduce has been experimented with as a general parallel approach for distributed computation of the minimal model and of the well-founded semantics, using an approach based on distributed graph computation~\cite{DBLP:conf/cilc/BortoliITDP19}.
However, the algorithms proposed are general,  architecture-independent and do not exploit any possible optimization. Thus, even if running time scales with the number of processors the performances are not  comparable to those of traditional methods in a single processor.


\section{Going Small: Logic Programming and GPUs}
\label{sec:gpus}
\emph{Graphical Processing Units (GPUs)} are massively parallel devices,
originally developed to efficiently implement graphics pipelines for the rendering of 2D and 3D scenes.
The use of GPUs has become pervasive in general-purpose applications that are not
directly related to computer graphics, but demand massive computational power
to process large amounts of data, such as molecular dynamics,
data mining, genome sequencing,  computational finance, etc.
Vendors such as AMD and NVIDIA provide dedicated APIs
and promote frameworks such as
\emph{OpenCL} \cite{opencl} and \emph{CUDA (Computing Unified Device Architecture)} \cite{CUDAZONE}
to support GPU-computing.
In this section we focus on the efforts  made to exploit  this type of parallelism in logic programming.

\subsection{GPU-Based Parallelism}

GPUs are designed to execute a very large number of concurrent threads on multiple data.
The underlying conceptual parallel model is defined as \emph{Single-Instruction Multiple-Thread (SIMT)}.
In the CUDA framework, threads are organized and executed in groups of 32 threads called \emph{warps}.
 Cores are grouped in a collection of \emph{Streaming MultiProcessors (SMs)} of the same size
 and warps are scheduled and executed on the SMs.
 Threads in the same warp are expected (but not forced) to follow the same program address.
 Whenever two (or more) groups of threads belonging to the
 same warp fetch/execute different instructions, \emph{thread divergence} occurs.
 In this case the execution of the different groups is serialized and the overall performance decreases.

From the programmer's perspective, threads  are logically grouped in 3D blocks and blocks are organized in 3D grids.
Each thread in a grid executes an instance of the same \emph{kernel} (namely, a C/C++ or Fortran procedure).
A typical CUDA program 
includes parts meant for execution on the CPU (the \emph{host}) and
parts meant for parallel execution on the GPU (the \emph{device}).
The host program contains instructions for device data initialization, grids/blocks/threads configuration, kernel launch,
and retrieval of results.
GPUs also exhibit a hierarchical memory organization.
The threads in the same block share data using high-throughput on-chip shared memory organized in \emph{bank}s
of equal dimension.
Threads of different blocks can only share data through the off-chip global memory.

\begin{figure}[bt]
{\centerline{\includegraphics[width=0.43\linewidth]{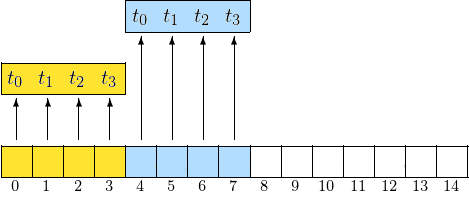}
~~~~~~
\includegraphics[width=0.43\linewidth]{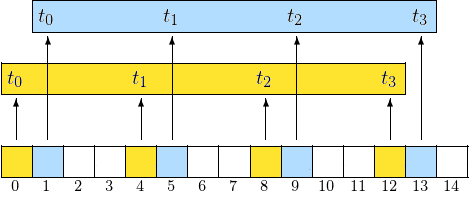}}
\caption{\label{fig:accesses}Memory-access patterns on an array data-structure by a group of four threads $t_0-t_3$.
Coalesced access pattern (left) and strided access pattern (right).}}
\end{figure}

To take full advantage of GPU architecture, one has to:
\begin{itemize}
\item
	proficiently distribute the workload among the cores to maximize GPU \emph{occupancy} (exploit all available
		device resources, such as SMs, registers, shared memory,...)
and minimize  \emph{thread divergence};
\item
achieve the highest possible throughput in memory accesses---namely, {\bf(1)}~adopt \emph{strided} access pattern to shared memory to minimize \emph{bank conflicts}
(i.e., accesses to locations in the same bank by threads of the same block. In this case accesses
		are serialized);
		{\bf(2)}~employ \emph{coalesced} accesses to global memory (see~Figure~\ref{fig:accesses}),
  as this minimizes the number of memory transactions.
\end{itemize}
These requirements  make the model of parallelization used on GPUs deeply different from those employed in
more ``conventional'' parallel architectures.
Existing serial or parallel solutions need substantial re-engineering to become profitably applicable in the context of GPUs.

\smallskip

The design of parallel engines for logic programming taking full advantage of
computational power of modern massively parallel graphic accelerators,
posed a number of challenges typically found in \emph{irregular applications}.
Briefly, applications are considered irregular when the exploitation of parallelism changes while
the execution proceeds.
Irregularity appears both in data accesses and in control flow. It is mainly due to the intrinsic
nature of the data, often represented through pointer-based data structures such as lists and graphs,
and to the concurrency patterns of the related algorithms.
Because of data-dependencies these applications frequently produce 
concurrent activity per data element, require unpredictable fine-grain communications, and 
exhibit peculiar load imbalances~\cite{lumsdaine2007challenges}.
Needless to say, the presence of irregularity makes it hard to maximize GPU occupancy and memory throughput,
while minimizing  thread divergence and bank conflicts.
This makes the development of solutions to irregular applications a difficult task,
where high performance and scalability are not an obvious outcome.
This is especially true when porting to GPUs  serial algorithms or even solutions originally
targeted at more traditional 
parallel/distributed architectures, such as cluster and multi-core systems.
Parallel graph algorithms constitute significant examples, that, like SAT/ASP solving,
are characterized by the need to process large, sparse, and unstructured data, and
exhibit irregular and low-arithmetic intensity combined with
data-dependent control flow and memory access patterns \cite{CUDASAT2,FGV17,DeclareYasmin19,FormisanoGV21}.

\subsection{GPU-Based  Datalog}

Datalog engines can be obtained by exploiting (parallel)
implementations of relational algebra operations \emph{select}, \emph{join}, and \emph{projection}. 
This approach is, in principle, viable also for GPU-based parallelism or, more generally, 
for the case of parallel/distributed heterogeneous architectures
(i.e., systems encompassing different devices, such as multi-cores, GPUs, FPGAs, etc).
Several proposals appeared in literature enabling the mechanization 
on (multi-)GPUs systems  and heterogeneous architectures,
of relational operators and SQL, also including aggregate operators 
\cite{007_HuangC15,033_WangXLSLG18,046_RuiT17,044_DiamosWWLY13,045_saeed15}. 

Concerning Datalog, Martinez-Angeles et al.~\citeyear{DUTRA13,082_Costa16} design a GPU-based engine by
exploiting GPU-accelerated relational algebra operators.
The computation order is driven by the dependency graph of the Datalog program and
fixpoint procedures are employed in case of recursive predicates.
The host preprocesses the program, converting each rule into an 
internal numerical representation;
the host decides which relational-algebra
operators are needed for each rule, while their executions are delegated to the device.
In particular, \emph{select} is implemented using three different device function executions.
The first one marks the rows of a matrix that satisfy the selection predicate, 
the second one performs a prefix sum on the marks to determine the
size of the results buffer and the location where each GPU thread must write
the results, and the last device function writes the results.
The \emph{projection} operator simply 
moves the elements of each required column to a different location. 
Concerning \emph{join}, the authors adopted a standard \emph{Indexed Nested Loop Join} algorithm.
We refer the reader to the mentioned works by Martinez-Angeles et al.~for the details on the CUDA implementation and
a report on experimental evaluations showing significant speedups of the GPU-based engine
with respect to engines running on single and multi-core CPUs. The experiments demonstrate scalability in presence
of extensional databases with several million of tuples. 

\smallskip

The Datalog-like language LogiQL is used as front-end in the Red~Fox system \cite{102_WuDSABGY14}.
LogiQL is a variant of Datalog including aggregations, arithmetic, integrity constraints and active rules.
Red Fox provides an environment enabling relational query processing on GPUs, through compilation
of LogiQL queries into optimized GPU-based relational operations. Then, an optimized query plan is generated and
executed on the device.
The approach is demonstrated to be faster than the corresponding (commercial) CPU-oriented implementation of LogiQL.

\smallskip

Nguyen et al.~\citeyear{084_NguyenSSI18} propose
a different approach to Datalog parallelization, not directly relying on parallelization of relational algebra.
In this work, the computation of the least model of a definite program
is obtained by first translating the program into a linear algebra problem.
Then, the multi-linear problem is solved on GPU by using standard CUDA libraries for matrix computations.
The solution of the multi-linear problem identifies the model of the program.
A similar approach is also viable to compute stable models of disjunctive logic programs. The approach demonstrates
encouraging performance results for randomly generated programs with over 20,000 rules.

\subsection{GPU-Based ASP}\label{sec:yasmin}
\SetKwFor{myFor}{for}{do}{}
\SetKwFor{myWhile}{while}{do}{}

\SetKwFunction{partitionproblem}{partition\_problem}
\SetKwFunction{yasminlauncher}{yasmin\_launcher}
\SetKwFunction{outputStableModels}{outputStableModels}
\SetKwFunction{pthreadcreate}{pthread\_create}
\SetKwFunction{pthreadjoin}{pthread\_join}
\SetKwFunction{pthreadexit}{pthread\_exit}
\SetKwFunction{yasminsingle}{yasmin}
\SetKwFunction{cudaStreamSynchronize}{cudaStreamSynchronize}
\SetKwFunction{cudaDeviceSynchronize}{cudaDeviceSynchronize}

The first attempt in exploiting GPU parallelism for ASP solving has been described
by Dovier et al.~\citeyear{DFPV15_ICLP,DFPV16,DeclareYasmin19},  proposing the solver {\sc yasmin}.
The authors design a conflict-driven ASP-solver
reminiscent of the conventional structure of sequential conflict-driven ASP solvers \cite{GebserKaminskiKaufmannSchaub12a}.
However, substantial differences lay in both the implemented algorithms and in the specific solutions
adopted to optimize GPU occupancy, minimize thread divergence, and maximizing memory throughput.
To this aim,  difficult to parallelize and intrinsically sequential components of serial solvers have
been replaced by parallel counterparts. More specifically, 
{\bf(1)}~the notion of \emph{ASP computations} \cite{ASPcomputation} is exploited to avoid the introduction
of loop formulas and the need of performing \emph{unfounded set checks}~\cite{GebserKaminskiKaufmannSchaub12a};
{\bf(2)}~a parallel conflict analysis procedure is used as an alternative to the sequential
resolution-based technique used in {\sc clasp}.
Memory accesses have been regularized
by suitably sorting input data with respect to size of rules,
by storing them using Compressed Sparse Row (CSR) format,
and by designing specific device functions, each one tailored to process groups of rules of homogeneous size. 
All this enables efficient balanced mapping of data to threads.
Moreover, to maximize device performance, the authors exploited specific
features supported by CUDA framework, such as 
\emph{shuffling} (a high efficient intra-warp communication mechanism) 
and
\emph{stream-based parallelism} (a support to concurrent kernels asynchronous execution available in the CUDA 
programming framework).

{\small
\begin{algorithm}[tb]
\DontPrintSemicolon
\KwInput{Ground ASP program $P$}
\KwInput{Number of subproblems $\mathit{Npb}$ and number of pthreads $\mathit{N}$}
\KwOutput{Stable models}
\CommentSty{\color{blue}}
	$\mathit{Parts} \gets \partitionproblem(\mathit{P}, \mathit{Npb}, \mathit{Opts})$ \label{A:linesplit}\\
                \tcc{spawns $\mathit{N}$ concurrent instances of $\yasminlauncher()$:}
	   \myFor{($\mathit{pth} \gets 0$; $\mathit{pth}<\mathit{N}$; $\mathit{pth}{+}{+}$)}{\label{A:line2}
                        $\mathit{pths}[\mathit{pth}] \gets \pthreadcreate(\yasminlauncher(\mathit{pth}))$ \label{A:line3}
        }
	   \myFor{($\mathit{pth} \gets 0$; $\mathit{pth}<\mathit{N}$; $\mathit{pth}{+}{+}$)}{\label{A:lineBarrier}
                        $\pthreadjoin(\mathit{pths}[\mathit{pth}])$\tcc*{wait for pthread's completion}
        }
	    \cudaDeviceSynchronize()\label{A:lineSync}\tcc*{wait for termination of all device code}
	    \outputStableModels()\label{A:lineOutput}
        \caption{\label{alg:A:multiyasmin}Host code of the main procedure {\sc pthreaded\_yasmin} of the pthreaded ASP-solver {\sc yasmin} ~~~\hfill(simplified)}
\end{algorithm}
}

GPU-based parallelism can be combined with host
parallelism by exploiting host POSIX \emph{pthreads} and CUDA streams.
In particular, Algorithm~\ref{alg:A:multiyasmin} (simplified form of the ASP-solver
designed by Dovier et al.~\citeyear{DeclareYasmin19})
shows the main part of the host code of the multi-pthread procedure  {\sc pthreaded\_yasmin}
which first splits (line~\ref{A:linesplit}) the given problem into a number $\mathit{Npb}$ of subproblems
 by applying some heuristics/criteria.
The simplest possibility would be to apply Or-parallelism and split the
search space by assigning in different ways the truth values of a subset of the input atoms.
Then, the procedure (lines~\ref{A:line2}--\ref{A:line3}) spawns a pool of $\mathit{N}$ 
host POSIX threads. 
Note that these pthreads share host variables (such as $P$, $\mathit{Parts}$,...),
but each pthread issues device commands in a different CUDA stream.
Each pthread created in line~\ref{A:line3} runs an instance of 
the host procedure \yasminlauncher\ (described in Algorithm~\ref{alg:B:multiyasmin}).
After the termination of all issued device commands (lines~\ref{A:lineBarrier}-\ref{A:lineSync})
results are collected and output (line~\ref{A:lineOutput}).

{\small
\begin{algorithm}[tb]
\DontPrintSemicolon
\KwInput{Pthread ID $\mathit{pth}$}
\KwOutput{Stable models (stored in global device memory)}
\CommentSty{\color{blue}}
        \myWhile(\tcc*[f]{repeat while there are subproblems}){\mbox{{\bf exists} } $S\in\mathit{Parts}$}
        {
         $\mathit{Parts} \gets \mathit{Parts}\setminus\{S\}$\label{B:line8}\\
	     $\yasminsingle(S,\mathit{pth})$\label{B:line9}\\
         \cudaStreamSynchronize()\tcc*[f]{wait for operations in the stream}
        }
        $\pthreadexit()$ \tcc*[f]{pthread ends and joins the main pthread}
	\caption{\label{alg:B:multiyasmin}Host code of {\sc yasmin\_launcher}, called in Algorithm~\ref{alg:A:multiyasmin}~~~\hfill(simplified)}
\end{algorithm}
}

Algorithm~\ref{alg:B:multiyasmin} describes the procedure \yasminlauncher\ executed
by each concurrent pthread. Such procedure iterates by
extracting (in mutual exclusion) one of the unsolved subproblems (line~\ref{B:line8}) and 
by running an instance of the CUDA solver in a dedicated CUDA stream (line~\ref{B:line9}).
Notice that, since each pthread runs an instance of the solver in a private CUDA stream,
each of them proceeds by issuing commands (memory transfers, kernel launches, etc)
in such stream, independently and concurrently with the other solver instances.
This helps in maximizing GPU occupancy, because it permits overlapping between
computation and memory transfers and allows scheduling of warps on all available SMs.

The combination of host-parallelism and device-parallelism opens up further refinements, such as 
the introduction of techniques like 
\emph{parallel lookahead} \cite{DBLP:books/sp/18/DovierFP18}, the development of more powerful 
\emph{multiple learning} schemes \cite{FVictcs14}, and paves the way to the exploitation of multi-GPU
and heterogeneous architectures in ASP-solving.

\section{Conclusion}
\label{sec:conclusion}

We have presented a review of the ``second twenty  years'' of 
research in parallelism and logic programming. 
The choice of the period is motivated by the availability of a
comprehensive survey of the first twenty years, 
published in 2001, that has served as a fundamental
reference to researchers and developers since.
While the contents of this classic survey are quite valid today, we
have attempted to gather herein the later evolution in the field,
which has continued at a fast pace, driven by the high speed of
technological evolution, that has led to innovations including very
large clusters, the wide diffusion of multi-core processors, the
game-changing role of general-purpose graphic processing units, or the
ubiquitous adoption of cloud computing.
In particular, after a quick review of the major milestones of the
first 20 years, we have reviewed the recent progress in parallel
execution of Prolog, including Or-parallelism, And-parallelism, static
analysis, and the combination with tabling. We have covered the
significant amount of work done in the context of parallelism and
Answer Set Programming and Datalog, including search parallelism, other forms of
parallelism, parallel grounding, or portfolio parallelism. Finally, we have 
addressed the connections with big data frameworks and graphical
processing units.


This new survey highlights once more the wide diversity of techniques
explored by the logic programming community to promote the efficient
exploitation of parallelism within the various variants of the
paradigm that have been emerging.
Over these years, we have seen the emergence of more declarative
styles of logic programming, such as Answer Set Programming, as well
as an evolution in this same direction within the other languages that
follow the Prolog-style line.
Nevertheless, the experiences reported in the survey 
show that many of the techniques developed in
the early days of parallel logic programming are still applicable and
have paved the way to the efficient parallelization of these more modern
logic programming variants. These lessons have also not been limited
to the domain of logic programming but have also benefited
exploitation of parallelism in other domains.
This includes, e.g., the static analysis examples in
Section~\ref{sec:staticanal} or 
the conceptual models used in or-parallelism, which have supported work on
parallel planning (e.g.,~\cite{parallel-plan1}). We expect the general
principles of parallel logic programming to remain valid and benefit
broader efforts to parallelism in even more domains.

Some of the works summarized in this survey are in their infancy, but they are expected
to become dominant trends in the years to come. Just as the use of GPUs has provided the backbone to success
of paradigms like machine learning, we expect GPUs to gain an even more prominent role in parallel 
logic programming---e.g., supporting more complex heuristics and novel extensions (such as the integration
of Answer Set Programming with constraint satisfaction and optimization). The role of multi-platforms is going
to become prominent, especially with the growing emphasis on edge-to-cloud computing.

We hope to have put together a worthy continuation of the classic survey,
covering these last twenty years, and hope that it will serve not only
as a reference for researchers and developers of logic programming
systems, but also as engaging reading for anyone interested in logic
and as a useful source for researchers in parallel systems outside
logic programming.
The interested reader will find details of the performance results obtained using a diversity of coding techniques, architectures and  benchmarks, 
in the original contributions cited in this paper.

\end{document}